\documentclass[twocolumn]{aastex631}

\usepackage{amsmath}
\usepackage[encapsulated]{CJK}

\newcommand{\Halpha}{H$\mathrm{\alpha}$} 

\newcommand{\prospector}{\texttt{Prospector}}

\begin{document}

\title{It's More Complicated Than You Think: A Forward Model to Infer the \\ Recent Star Formation History, Bursty or Not, of Galaxy Populations}

\correspondingauthor{Emilie Burnham}
\email{efb5552@psu.edu}

\author[0000-0001-8174-317X]{Emilie Burnham}
\affiliation{Department of Astronomy \& Astrophysics, The Pennsylvania State University, University Park, PA 16802, USA}
\affiliation{Institute for Gravitation and the Cosmos, The Pennsylvania State University, University Park, PA 16802, USA}
\affiliation{Center for Astrostatistics and Astroinformatics, The Pennsylvania State University, University Park, PA 16802, USA}

\author[0000-0001-9269-5046]{Bingjie Wang (\begin{CJK*}{UTF8}{gbsn}王冰洁\ignorespacesafterend\end{CJK*})}
\thanks{NHFP Hubble Fellow}
\affiliation{Department of Astrophysical Sciences, Princeton University, Princeton, NJ 08544, USA}

\author[0000-0001-6755-1315]{Joel Leja}
\affiliation{Department of Astronomy \& Astrophysics, The Pennsylvania State University, University Park, PA 16802, USA}
\affiliation{Institute for Computational \& Data Sciences, The Pennsylvania State University, University Park, PA 16802, USA}
\affiliation{Institute for Gravitation and the Cosmos, The Pennsylvania State University, University Park, PA 16802, USA}

\author{Owen Gonzales}
\affiliation{Department of Astronomy, Van Vleck Observatory, Wesleyan University, Middletown, CT 06459, USA}

\author[0000-0002-5612-3427]{Jenny E. Greene}
\affiliation{Department of Astrophysical Sciences, Princeton University, Princeton, NJ 08544, USA}

\author[0000-0001-9298-3523]{Kartheik G. Iyer}
\affiliation{Columbia Astrophysics Laboratory, Columbia University, New York, NY 10027, USA}
\affiliation{Center for Computational Astrophysics, Flatiron Institute, New York, NY 10010, USA}

\author[0000-0002-9816-9300]{Abby Mintz}
\affiliation{Department of Astrophysical Sciences, Princeton University, Princeton, NJ 08544, USA}

\author[0000-0003-4075-7393]{David J. Setton}
\thanks{Brinson Prize Fellow}
\affiliation{Department of Astrophysical Sciences, Princeton University, Princeton, NJ 08544, USA}

\author[0000-0002-3977-2724]{Sarah Wellons}
\affiliation{Department of Astronomy, Van Vleck Observatory, Wesleyan University, Middletown, CT 06459, USA}

\author[0000-0001-5063-8254]{Rachel Bezanson}
\affiliation{Department of Physics and Astronomy and PITT PACC, University of Pittsburgh, Pittsburgh, PA 15260, USA}

\author[0000-0002-0212-4563]{Olivia Curtis}
\affiliation{Department of Astronomy \& Astrophysics, The Pennsylvania State University, University Park, PA 16802, USA}
\affiliation{Institute for Gravitation and the Cosmos, The Pennsylvania State University, University Park, PA 16802, USA}
\affiliation{Penn State Extraterrestrial Intelligence Center, The Pennsylvania State University, University Park, PA 16802, USA}

\author[0000-0002-1109-1919]{Robert Feldmann}
\affiliation{Department of Astrophysics, Universität Zürich, Winterthurerstrasse 190, Zürich, CH-8057, Switzerland}

\author[0000-0001-8367-6265]{Tim B. Miller}
\affiliation{Center for Interdisciplinary Exploration and Research in Astrophysics (CIERA), Evanston, IL 60201, USA}

\author[0000-0003-2804-0648 ]{Themiya Nanayakkara}
\affiliation{Centre for Astrophysics and Supercomputing, Swinburne University of Technology, Hawthorn, VIC 3122, Australia}

\author[0000-0003-2573-9832]{Joshua S. Speagle (\begin{CJK*}{UTF8}{gbsn}沈佳士\ignorespacesafterend\end{CJK*})}
\affiliation{David A. Dunlap Department of Astronomy and Astrophysics, University of Toronto, Toronto, Ontario, M5S 3H4, Canada.}
\affiliation{Dunlap Institute for Astronomy and Astrophysics, Toronto, Ontario M5S 3H4, Canada}
\affiliation{Department of Statistical Sciences, University of Toronto, Toronto, ON, M5G 1Z5, Canada}
\affiliation{Data Sciences Institute, University of Toronto, Toronto, ON, M5G 1Z5, Canada.}

\author[0000-0002-1714-1905]{Katherine A. Suess}
\affiliation{Department for Astrophysical \& Planetary Science, University of Colorado, Boulder, CO 80309, USA}

\author[0000-0003-4070-497X]{Guochao Sun}
\affiliation{Center for Interdisciplinary Exploration and Research in Astrophysics (CIERA), Evanston, IL 60201, USA}

\begin{abstract}
Observations of the early Universe (z $\gtrsim$ 4) with the James Webb Space Telescope reveal galaxy populations with a wide range of intrinsic luminosities and colors. Bursty star formation histories (SFHs), characterized by short-term fluctuations in the star formation rate (SFR), may explain this diversity, but constraining burst timescales and amplitudes in individual galaxies is challenging due to degeneracies and sensitivity limits. We introduce a population-level simulation-based inference framework that recovers the power and timescales of SFR fluctuations by forward-modeling galaxy populations and distributions of rest-UV to rest-optical spectral features sensitive to star formation timescales.
We adopt a stochastic SFH model based on a power spectral density formalism spanning 1 Myr-10 Gyr. Using simulated samples of $N=500$ galaxies at $z\sim4$ with typical JWST/NIRSpec uncertainties, we demonstrate that:
(i) the power of SFR fluctuations can be measured with sufficient precision to distinguish between simulations (e.g., FIRE-2-like vs. Illustris-like populations at $>$99\% confidence for timescales $<$ 100 Myr);
(ii) simultaneously modeling stochastic fluctuations and the recent ($t_L <$ 500 Myr) average SFH slope is essential, as secular trends otherwise mimic burstiness in common diagnostics;
(iii) frequent, intense bursts impose an outshining limit, and bias inference toward underestimating burstiness due to the obscuration of long-timescale power; and
(iv) the power of SFR fluctuations can be inferred to 95\% confidence across all timescales in both smooth and bursty populations.
This framework establishes a novel and robust method for placing quantitative constraints on the feedback physics regulating star formation using large, uniformly selected spectroscopic samples.
\end{abstract}


\section{Introduction} \label{sec:intro}

The observed diversity of galaxies is thought to arise from the interplay of physical processes operating over a wide range of timescales, from the rapid collapse of molecular clouds on timescales of $\sim$ 1-10 Myr to the long-term effects of stellar feedback, gas accretion, and environment up to 10 Gyr timescales (see \citealt{McKeeOstriker2007} and \citealt{Iyer2020} for reviews). A central goal of galaxy evolution studies is therefore to recover the star formation histories (SFHs) that are the direct results of these processes. Interpreting the spectral energy distributions (SEDs) of galaxies holds the promise of inferring accurate and well-calibrated SFHs, but doing so requires assumptions about how star formation proceeds over time. Although SFHs may be arbitrarily complex, they have long been approximated using smooth parametric forms. While such assumptions were necessary at the time due to limited wavelength coverage and data quality, such models can obscure the fact that star formation is regulated by additional physical mechanisms, particularly in low-mass galaxies \citep{Weisz2012}. A growing body of evidence instead indicates that galaxies undergo bursts of star formation, and that these bursts can significantly alter the physical properties inferred from their SEDs \citep{Johnson2013,Sparre2017,Caplar2019,Faisst2019,Emami2019,Johnson2020,WangLilly2020a,FloresVelazquez2021,Narayanan2023,Haskell2024,Sun2025,Cole2025,Munoz2026}. Understanding how bursts can be robustly inferred from galaxy SEDs, especially in the early Universe, is therefore essential for constraining galaxy formation timescales and evolutionary pathways.

The James Webb Space Telescope (JWST) has ushered in a new era of discovery in the early Universe by probing galaxy populations at high redshifts with both photometric \citep[e.g.,][]{Bezanson2024, Treu2022} and spectroscopic surveys \citep[e.g.,][]{Eisenstein2023,deGraaff2024}. Early investigations of the high-$z$ ultraviolet (UV) luminosity function revealed an apparent excess of UV-luminous galaxies compared to pre-JWST expectations \citep[e.g.,][]{Harikane2023, Mason2023}. In addition, galaxies exhibiting Balmer breaks at $z>5$ have been identified \citep[e.g.,][]{Looser2024}, and sometimes along with emission lines \citep{Witten2025}, indicating recent or ongoing (mini-)quenching events. 

This observed diversity of inferred intrinsic galaxy properties is highly puzzling.
A variety of physical processes have been proposed to explain this discrepancy between expectations and observations, which include a non-universal or top-heavy initial mass function (IMF) resulting in an abundance of rest-frame UV radiation massive stars \citep{Finkelstein2023, Harikane2023a, Munoz2023}, little to no suppression of star-formation from the UV background \citep{Dekel2023, Harikane2023a, Mason2023,Somerville2015}, or presence of active galactic nuclei \citep[AGN;][]{Ono2018,Pacucci2022}.

A further compelling argument is the presence of large-amplitude, short-timescale fluctuations of the SFR, referred to as “bursty’’ SFHs \citep[e.g.,][]{Mirocha2023,Mason2023,Pallottini2023,Sun2023:bright_galaxies,Munoz2023,Shen2023}. In the local Universe, it is understood that the timescale for star formation stems from the quasi-equilibrium of feedback from young, massive stars and supernovae with the accretion timescale of cold gas. This results in star formation that proceeds smoothly when averaged over long timescales in massive ($M_\star\gtrsim10^{10}$ M$_\odot$) galaxies \citep{FaucherGiguere2018}. Bursty star formation is characterized by the feedback timescale being comparable to or less than the timescale of the dynamical time for star formation. This feedback can delay star formation by preventing gas from cooling and collapsing \citep{FaucherGiguere2018,Tacchella2022}. As young, massive stars evolve off the main sequence, ionizing radiation drops. This allows nearby gas to recombine, cool, and form a new stellar population. In bursty populations, shallow surveys primarily detect galaxies in their burst phase, while those in ``mini-quench'' or ``napping'' phases remain relatively hidden \citep{Strait2023,Sun2023,Looser2024,Wang2025:sfh,Endsley2025}. This notion can account for both the UV-luminous and mini-quenched observed galaxies at high redshifts with JWST.

Indeed, the Feedback In Realistic Environments (FIRE) cosmological zoom-in simulations \citep{Hopkins2014:fire,Hopkins2018:fire2} have demonstrated bursty SFHs across a wide range of galaxy masses at $z > 1$, where stellar feedback is resolved both spatially and temporally \citep{FaucherGiguere2018}. Bursty star formation is not only a natural outcome of resolved feedback, but is also required in simulations to reproduce a variety of observables, including solving the core-cusp problem in low-mass galaxies \citep{Brooks2013,Somerville2015}, matching the low-mass stellar-halo mass relation \citep{Behroozi2019}, and explaining low star formation efficiencies in local galaxies \citep{Semenov2021}. Although often calibrated to reproduce observations, modern cosmological simulations differ in their treatment of sub-grid physics. This results in variations in the timescales governing galaxy growth and star formation that depend on the specific feedback prescriptions \citep{Somerville2015,Naab2017,Dome2023}. \citet{Iyer2020} investigated the variability of an extensive set of simulated SFHs from cosmological hydrodynamical simulations, semi-analytic, and empirical models. The authors found a considerable diversity in the power of said fluctuations at a given timescale, the amount of correlation between timescales, and the evolution with respect to stellar mass across the set of models investigated. These discrepancies underscore the need for observational constraints on short-term SFR fluctuations to falsify differing feedback models between cosmological simulations, especially because many simulations can be calibrated to reproduce qualitatively similar cosmic SFR densities and galaxy stellar mass functions, which alone do not allow us to discriminate between feedback prescriptions. With recent improvements in spatial and temporal resolution, simulations are now able to capture these rapid fluctuations, providing a useful theoretical backdrop for interpreting observational measurements (see \citealt{Vogelsberger2020} for a review).

While simulations highlight the diversity of SFHs, translating these insights into constraints from observed galaxy populations remains challenging. Outshining poses a fundamental challenge for reconstructing SFHs from individual galaxies; very young stellar populations dominate the integrated light \citep[e.g.,][]{Papovich2001}, suppressing the observable signatures of older ($\gtrsim100$ Myr) populations whose spectral features evolve slowly \citep{Ocvirk2006}. Consequently, rising or bursty SFHs can produce similar SEDs when recently formed ($\lesssim10$ Myr) populations dominate, making it difficult to recover the underlying SFH from individual galaxies. Indeed, \citet{Wang2025:sfh} demonstrated that even with high-quality spectra, fitting galaxies individually often fails to recover complex, rapidly varying SFHs because the SED of a single galaxy lacks sufficient information.

These challenges motivate the use of population-level approaches, which leverage the distribution of timescale-sensitive features across many galaxies. A widely used example is the H$\alpha$/UV ratio, which can be used as a population-level burstiness diagnostic \citep[e.g.,][]{Weisz2012,Johnson2013,Broussard2019,Faisst2019,Emami2019,FloresVelazquez2021,Asada2024}. Sensitive to fluctuations on $\sim$10-70 Myr timescales, H$\alpha$/UV can identify recent bursts of star formation. However, it cannot capture the full complexity of galaxy SFHs \citep{Mehta2023, Wang2025:sfh}. The ratio responds not only to burstiness but also to smooth long-term trends in the SFH, and is further influenced by metallicity and dust attenuation at levels comparable to burstiness. Because H$\alpha$ and UV probe partially overlapping stellar ages, the ratio provides only weak constraints on burst timescales, highlighting the need for multi-timescale indicators that can probe both short- and long-term SFR fluctuations.

More comprehensive population-level methods forward-model galaxy populations and compare the simulated distributions of timescale-sensitive features to observations. These features trace different timescales because they respond to stellar populations of different characteristic lifetimes: rest-optical emission lines are powered by ionizing photons from short-lived, massive O-B type stars, while continuum features, such as the Balmer break, is the result of the accumulated light of longer-lived A-F type stars \citep{Murphy2011,Hao2011}. Recent work demonstrates that this approach can accurately recover key SFH characteristics. \citet{Iyer2025} find that samples as small as $\sim30$ galaxies can constrain typical burstiness, and \citet{Wang2025:sfh} show that the amplitude, duration, and slope of recent variability are encoded in the population-level distributions. This framework has already proven effective in observational studies of SFH burstiness and duty cycles \citep[e.g.,][]{Lee2009,Papovich2011,Jaacks2012,Weisz2012,Emami2019,Atek2022,Cole2025,Mintz2025} and provides a foundation for improving the inference of individual galaxy SFHs and physical properties in future analyses.

Building upon these population-level approaches, a full forward-model inference framework would allow direct constraints on the posterior distributions of burstiness parameters for a chosen SFH parameterization. This contrasts with inferring burstiness indirectly from the overall scatter or distribution of timescale-sensitive features. Such an approach would forward-model arbitrarily complex SFHs while simultaneously predicting a wide set of timescale-sensitive observables, enabling quantitative constraints on burstiness that are only weakly dependent on the assumed SFH parameterization. Historically, this was limited by the computational cost of stellar population synthesis (SPS) modeling (e.g., \texttt{FSPS}; \citealt{Conroy2009, Conroy2010}). Each model evaluation requires $\sim$10-100 ms, making large-scale population-level inference impractical. Even neural-network emulators of SPS outputs \citep[e.g.,][]{Alsing2020, Mathews2023}, reduce this cost, but no existing framework can efficiently handle arbitrarily complex, fine time-resolution SFHs. 

Simulation-based inference (SBI) overcomes these limitations by directly learning a neural approximation to the posterior from forward-modeled simulations \citep[see review by][]{Cranmer2020}. SBI is especially powerful when an analytical likelihood function is intractable or too expensive to compute. In this context, the black-box simulator comprises the assumed SFH population model, stochastic realizations of individual SFHs, the SPS framework, and observational uncertainty, allowing SBI to naturally marginalize over stochasticity and nuisance parameters while conditioning on the population-level burstiness parameters of interest. Importantly, utilizing SBI reduces full population-level fits of 500 galaxies from tens of thousands of computing hours to mere seconds.

In this work, we present an inference framework that leverages SBI to forward-model galaxy populations from a specified population model. To demonstrate the methodology, we consider two population SFH models: (1) an illustrative oscillating SFH characterized by burst timescales, amplitudes, and slopes, and (2) a stochastic model that generates correlated, sloped SFHs from a power spectral density (PSD). The single-frequency model, used as a proof of concept, accurately recovers population-level burst parameters, validating the robustness of the framework. The PSD formalism, which captures stochastic SFH variability across 1 Myr-10 Gyr timescales, shows considerable promise for constraining the underlying SFH PSD from observed galaxy populations and for observationally probing feedback prescriptions in cosmological simulations.

This paper is organized as follows. Section \ref{sec:SFHModels} introduces the population SFH models considered in this work. Section \ref{sec:methods_population_modeling} describes our approach to modeling galaxy populations and inferring SFH timescales, and Section \ref{sec:methods_sbi} presents the SBI inference framework. Section \ref{sec:Results} outlines how the distribution of observed SFR timescale indicators varies with the assumed population model and evaluates the ability of the SBI framework to recover population-level SFH parameters for both models. We discuss the broader implications of these results in Section \ref{sec:discussion}, and we summarize our conclusions in Section \ref{sec:conclusions}. We adopt the best-fit cosmological parameters from the Wilkinson Microwave Anisotropy Probe 9-year mission: $H_0 = 69.32 \ \mathrm{km \ s^{-1} \ Mpc^{-1}}$, $\Omega_M = 0.2865$, and $\Omega_{\Lambda} = 0.7135$ \citep{Bennett2013}. 

\section{Population Models of Galaxy Star Formation Histories}\label{sec:SFHModels}

In this section, we introduce the formalism underlying the two population-level SFH models utilized in this work. The first is a single-frequency toy model that provides an illustrative baseline for the proposed forward model. The second is a more physically motivated, PSD-based model designed to capture the richer, multi-frequency variability expected in the SFHs of real galaxy populations. These models are intentionally distinct: the first serves as a proof-of-concept, while the second constitutes the forward model intended for application to real galaxy populations.

\subsection{A Single-Frequency SFH Population Model}\label{sec:simple_model}

\begin{figure*}
    \centering
    \includegraphics[width=1\linewidth]{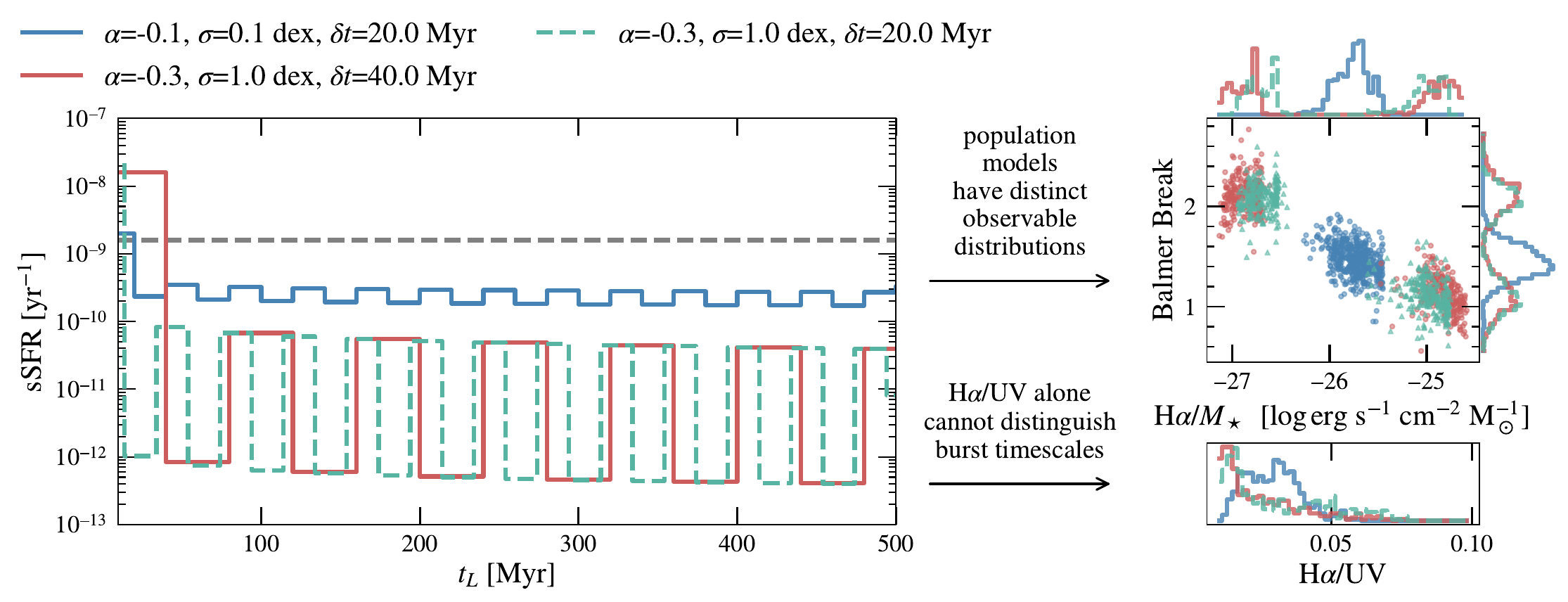}
    \caption{Effect of the single-frequency illustrative model on SFR indicator distributions. Left panel: Three SFH model realizations, with $\alpha=0.1$, $\sigma=-0.1$, and $\delta t = 20$ Myr (blue), $\alpha=-0.3$, $\sigma=1.0$, and $\delta t = 40$ Myr (red), and $\alpha=-0.3$, $\sigma=1.0$, and $\delta t = 20$ Myr (teal dashed) in a burst phase. The gray dashed line indicates the SFMS at $z=4$ to be sSFR$_\text{SFMS} = 10^{-8.8} \ \mathrm{yr^{-1}}$. Right panel: The marginal and joint distribution of stellar mass normalized H$\alpha$ flux and the Balmer break strength for 500 samples of each SFH model realization. We marginalize over stellar metallicity and the SFH phase for each sample. The distribution of H$\alpha$/UV flux from each population model is shown in the bottom right panel.}
    \label{fig:simple_model}
\end{figure*}

To investigate the efficacy of our method for inferring the timescales of star formation in galaxy populations, we utilize an illustrative single-frequency oscillating SFH model\footnote{The single-frequency SFH is implemented as a square-wave function with an added slope. While the PSD of a square wave formally reports power at multiple harmonics, we refer to it as ``single-frequency” to indicate that there is a single characteristic burst timescale of interest, and to facilitate comparison with the PSD formalism.}. This is identical to the SFH model introduced in \citet{Wang2025:sfh}. The single-frequency model is designed to capture a few key features of the diversity in galaxy populations at high redshifts, specifically the range in observed SFRs and non-constant SFHs. The model has four parameters:

\begin{enumerate}
    \item $\sigma$\footnote{We denote the difference between the dispersion of a sample, or the width of a normal distribution, as $1\sigma = (P_{84} - P_{16})/2$, and the single-frequency SFH oscillation amplitude parameter as $\sigma$.}: The amplitude of the square-wave oscillations in sSFR around the star-forming main sequence (SFMS); measured in dex. Values of $\sim$0.3 dex correspond to fluctuations comparable to the observed $1\sigma$ scatter of the local SFMS \citep{Speagle2014, Leja2022}, representing relatively “normal” variability, whereas values $\gtrsim 0.5$ dex correspond to more “bursty” behavior \citep{Hopkins2018:fire2}. Here, $\sigma$ parametrizes the instantaneous deviation of the current sSFR with respect to the SFMS; it is not a time-averaged quantity.

    \item $\delta t$: The timescale for oscillations in star formation ($1/2$ period of the oscillation); measured in Myr of look-back time. A timescale of $\delta t \approx 40$ Myr approximates the simulated burst duty cycle in the FIREbox SFHs \citep{Feldmann2023}.
    
    \item $\alpha$: The power-law index that describes the slope of the rising SFH over the last 500 Myr, such that $\text{SFR}(t_L \leq 500 \text{ Myr}) \propto t_L^\alpha$. A shallowly rising SFH corresponds to $\alpha \approx -0.1$, whereas a steeply rising SFH has $\alpha \approx -0.3$.
    
    \item $\phi$: The oscillation phase of the currently observed SFR, with $\phi \in [0,2\pi)$. When $\phi < \pi$, the galaxy is in a ``burst" phase, and when $\phi \geq \pi$, the galaxy is in a ``mini-quench" phase. Sampling over $\phi$ simulates observing a population of galaxies at different phases of their SFHs.

\end{enumerate}

\noindent The left panel of Figure \ref{fig:simple_model} displays three distinct single-frequency SFH model realizations. Here, we assume that all SFHs rise to the SFMS at $z=4$ with some power-law slope $\alpha$, which is indicated by the gray dashed line at sSFR$_{\text{SFMS}} = 10^{-8.8} \ \mathrm{yr^{-1}}$. 

\subsection{A Flexible Stochastic SFH Population Model}\label{sec:psd_model}

\begin{figure*}
    \centering
    \includegraphics[width=1\linewidth]{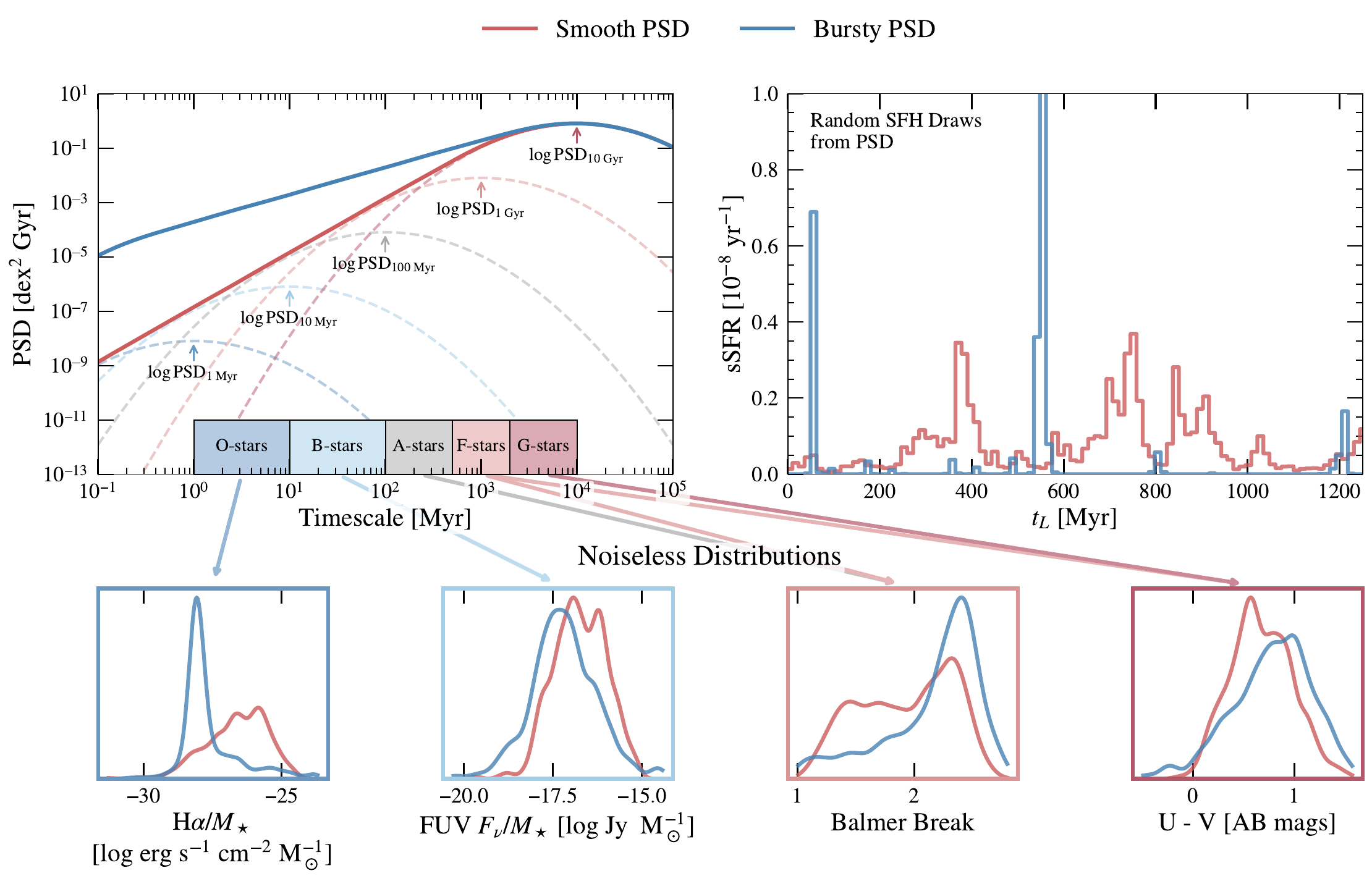}
    \caption{Effect of the flex-PSD model on observable distributions. Upper left panel: Two realizations of the flex-PSD model, a bursty model (blue, power-law slope of $\beta = 1$), and a smooth model (red, power-law slope of $\beta= 2$). For the smooth model, we display the log-normal Gaussian components as dashed lines with each normalization annotated with the flex-PSD model parameter name. We additionally show the approximate timescales that are probed by the dominant stellar spectral type in the integrated light of an SED: O stars on 1-10 Myr timescales, B stars on 10-100 Myr timescales, A stars on 100-500 Myr timescales, F stars on 0.5-2 Gyr, and G stars $> 2$ Gyr. Upper Right Panel: Two example SFHs sampled from the two models. Lower panels: Marginal distributions of observed spectral features from mock populations simulated from each PSD. We include stellar-mass normalized H$\alpha$ flux, stellar-mass normalized FUV flux density, Balmer break strength, and rest-frame U-V color.}
    \label{fig:flexpsd_model}
\end{figure*}

Real galaxies can be more complex, leading to SFHs that are not well captured by the single-frequency model. To address this, we introduce a population model capable of generating multi-frequency correlated SFHs via a PSD. Cosmological simulations show that different subgrid feedback prescriptions produce clearly distinguishable SFH PSDs \citep{Iyer2020}, demonstrating that the PSD encodes key physical differences in galaxy formation. PSDs have also been used as a framework for modeling SFHs in a variety of contexts \citep[e.g.,][]{Kravtsov2024,Sun2025,Carvajal-Bohorquez2025}, and any given PSD can serve as a prior on the SFH of a galaxy in an individual SED fit \citep{Wan2024,Wan2025}.

For a continuous time series, where the SFH = SFR($t$), the PSD can be calculated by the Fourier transform of the signal at a given timescale, $\tau = t' - t$:

\begin{equation}\label{eq:PSD}
    \text{PSD}(\tau) = \left| \int_{-\infty}^{\infty} \log\text{SFR}(t)e^{-2\pi i\tau t} dt \right|^2
\end{equation}

In this work, we adopt a flexible PSD parametrization capable of describing both extremely bursty and smooth SFHs. We model the PSD as a sum of five log-normal Gaussian components, with means $\mu = [-3,-2,-1,0,1]$ logGyr and widths $\sigma = 0.5$ logGyr, ensuring a smooth PSD shape with coverage of timescales from 1 Myr to the age of the Universe. For illustrative purposes, we denote the log-normalization of each component as logPSD$_i$, where $i$ labels the characteristic timescale of the component: 1 Myr, 10 Myr, 100 Myr, 1 Gyr, or 10 Gyr. The formalism is therefore described by

\begin{multline}
        \text{PSD}_{\text{flex}}(\tau\text{ [Gyr]}) = \frac{1}{\sigma\sqrt{2\pi}}\sum_i 10^{\log \text{PSD}_i} \times \ \\  \exp\left[\left(\frac{\log{\tau} - \mu_i}{\sigma \sqrt{2}}\right)^2\right] [\mathrm{dex^2 \ Gyr}].
\end{multline}

\noindent Hereafter, this SFH model is referred to as the ``flex-PSD" model. The parametrization of the flex-PSD model is depicted in the top-left panel of Figure \ref{fig:flexpsd_model}. Two distinctive PSDs are presented: a bursty model, characterized by a shallow PSD (blue), and a smooth model, characterized by a steep PSD (red). These PSDs generate significantly different SFHs, as evident in the top right panel of the figure. 

When Equation \ref{eq:PSD} is calculated, all phase information of the original SFH is lost. As in, any rising or falling of the SFH at the time of observation is not retained throughout the method, as the PSD only reports the power at that specific mode. To counteract this, we include two additional parameters to model the slope of the recent ($t_L\leq500$ Myr) SFH. These parameters are as follows:

\begin{enumerate}
\item $\mu_\alpha$:  Mean power-law slope of a population's SFHs over the past 500 Myr in look-back time such that $\left<\text{SFR}(t_L \leq 500 \text{Myr})\right>_\text{pop} \propto t_L^{\mu_\alpha}$. For example, a value of $\mu_\alpha = -1$ indicates that the mean SFH in a population is rising (SFR increasing on average), where $\mu_\alpha = 1$ is when the mean SFH is falling (SFR decreasing on average).
    \item $\sigma_\alpha$: Dispersion of the power-law slopes of a population's SFHs, assuming a normal distribution. For a given sampled galaxy from a population, the power-law slope, $\alpha$ (identical to the $\alpha$ parameter in the single-frequency SFH model), is sampled from a normal distribution such that $\alpha \sim \mathcal{N}(\mu_\alpha,\sigma_\alpha)$. 
\end{enumerate}

The flex-PSD framework assumes that SFHs are approximately stationary over the timescales probed by the PSD. Real galaxies often exhibit secular trends, such as long-term rises or declines in SFR, which formally violate this assumption; our slope parameters ($\mu_\alpha$, $\sigma_\alpha$) help account for these population-level non-stationary signals in galaxy SFHs. Additionally, because the PSD is defined in terms of $\log\text{SFR}$, the model implicitly assumes that deviations from the mean SFH are roughly log-normally distributed. Therefore, extreme, non-Gaussian bursts may be underrepresented in the inferred PSDs. Nevertheless, these limitations are not expected to materially affect the primary results, and the framework remains robust for characterizing the overall burstiness of galaxy populations.

\section{Modeling Observables from Mock Populations}\label{sec:methods_population_modeling}

In this section, we describe how SFR timescale indicators are observed and modeled, and how the distribution of these spectral features across a galaxy population can directly constrain the population's burstiness.

\subsection{SFR Timescale Indicators}\label{sec:sfrt_indicators}

Galaxy spectra contain features that correlate with SFRs over different timescales \citep[see review by][]{Kennicutt2012}. In this work, we focus on features that probe a wide range of timescales, listed below. We note that these timescales are not fixed: the effective timescale that a given feature probes depends on the underlying SFH, particularly in galaxies with steeply rising or falling star formation. This dependence motivates our full forward-modeling approach, which accounts for how the diversity of SFHs in a population maps onto the distribution of observable spectral features.

We consider the following spectral features, loosely following \citet{Iyer2024, Wang2025:sfh}:

\begin{enumerate}
    \item Balmer emission line flux: 
    Balmer line emission originates in HII regions, where hydrogen is ionized by UV radiation from young massive stars (primarily O- and early B-type stars). Strong Balmer emission is a robust indicator of very recent star formation, probing timescales of roughly $<10$ Myr. In this work, we focus on the \Halpha\ line due to its prominence as the strongest rest-optical hydrogen emission line.\footnote{From an observational perspective, equivalent widths (EWs) are generally preferred for measuring emission line strength because they are continuum-normalized, and thus independent of stellar mass. However, since we work purely in model space, the line flux and EW contain the same information content.}

    \item Dust-corrected rest-frame UV flux densities:
    UV radiation is predominantly emitted by hot, massive stars with effective temperatures $T_{\text{eff}} \gtrsim 10,000$ K. These stars have lifetimes $\lesssim 100$ Myr, with the far-UV (FUV; $\lambda \in [1000,1700] \ \mathrm{\AA}$) dominated by the shortest-lived O- and B-type stars and the near-UV (NUV; $\lambda \in [1700,3200] \ \mathrm{\AA}$) including contributions from slightly longer-lived early- to mid-B type stars. Therefore, UV flux densities trace star formation on timescales of roughly $10 - 100$ Myr. FUV is more sensitive to the very recent star formation ($\sim 10-30$ Myr), while NUV captures somewhat longer timescales ($\sim 30-100$ Myr).

    \item Balmer break strength: 
    The Balmer break arises due to bound-free absorption from hydrogen in the $n=2$ energy level, producing a spectral discontinuity in A-type and late-B type stars \citep[e.g.,][]{Suess2022}. Balmer breaks are strongest in A-type stars, making them tracers of star formation on $\sim 100$ Myr timescales. Unlike the commonly used $D_{4000}$ index \citep{Bruzual1983,Balogh1999}, which measures the 4000 \AA\ break caused by metal absorption features in G-type stars and cooler, we adopt the modified Balmer break definition proposed by \citet{Wang2024:rubies}, optimized for high-redshift galaxies where the 4000 \AA\ break is weak due to the age of the Universe. 

    The Balmer break strength in this analysis is defined as:
\end{enumerate}
\begin{equation}\label{eq:balmerbreak}
 \frac{\text{Median}(F_{\nu}, \lambda \in [4000, 4100] \, \text{\AA})}{\text{Median}(F_{\nu}, \lambda \in [3620, 3720] \, \text{\AA})}.
\end{equation}

\begin{enumerate}
    \item[] For a detailed discussion of this approach, we refer the reader to Appendix A of \citet{Wang2024:rubies}.

    \item[4.] U-V rest-frame color: 
    The rest-frame U-V color probes the optical continuum, comparing light from young, massive stars to that from older, intermediate-mass stars. A red (more positive) U-V color indicates either an old stellar population or significant dust attenuation. U-V, especially when combined with V-J, is commonly used to classify galaxies as star-forming or quiescent. As we will illustrate shortly, the dust-free rest U-V color primarily traces star formation over longer timescales, $\gtrsim 1$ Gyr, corresponding to the lifetimes of low- to intermediate-mass stars dominating the optical continuum.
\end{enumerate}

Together, these indicators provide a multi-timescale, from $1$ Myr to $>$ 1 Gyr, view of star formation activity in galaxies. We next discuss how they are modeled with SPS methods to produce mock observables for galaxy populations.

\subsection{Simulating Observables from Galaxy Populations}\label{sec:methods_sps}

We simulate galaxy observables using \texttt{FSPS} \citep{Johnson2024_pythonfsps} accessed via \texttt{python-fsps}, in combination with \texttt{Prospector} \citep{Johnson2021_prospect}. Stellar population spectra are computed using the MIST isochrones \citep{Choi2016, Dotter2016} and the MILES spectral library \citep{SanchezBlazquez2006}. Nebular emission is included via a pre-computed \texttt{Cloudy} \citep{Ferland2017} emission line grid \citep{Byler2017}, with the ionization parameter and gas-phase metallicity fixed to $\log U = -1$ and $\log Z_{\rm gas}/Z_\odot = -0.5$, respectively. We discuss the potential systematics introduced by these SPS choices in Section \ref{sec:discussion_systematics}.

We assume no internal dust attenuation within the model galaxies. In applications to observed populations, per-galaxy dust corrections can be applied independently, either using the Balmer decrement ($F_{\rm H\alpha}/F_{\rm H\beta}$) or by inferring a dust model conditioned on the observed photometry. We discuss the implications of this choice in Section \ref{sec:discussion_systematics}. All galaxies are modeled at $z = 4$ with stellar masses of $M_\star = 10^{10} \ \mathrm{M_\odot}$. This choice of stellar mass is entirely arbitrary, as the observables we focus on are normalized by the stellar mass and the SFH models do not depend on it. This redshift is chosen so that the H$\alpha$ line falls within JWST's near-infrared spectrograph observed wavelength window. Note that our conclusions are generalizable to a population of any stellar mass; the most relevant parameter for these observables is redshift, as it determines the age of the universe and thus the relative strength of old stars, which contribute at a low level to the timescale-sensitive spectral features. 

Stellar metallicity is a dominant source of systematic uncertainty in SFR-sensitive features: even modest metallicity variations can alter line strengths and continuum shapes on the order of a factor of $2\times$, introducing covariances that could be misattributed to burstiness \citep{Bicker2005,Leja2017,Park2025:alpha}. To account for this, we marginalize over a broad, uniform distribution of stellar metallicities, $\mathcal{U}(-2.00,0.19) \ \log Z_\star / Z_\odot$, when generating our model training set. This approach assumes a large intrinsic spread in metallicity within each galaxy population, making it a conservative choice. It ensures that the resulting SBI model is insensitive to metallicity-driven dispersion in the features and instead learns how observable distributions respond primarily to differences in the underlying population SFH. We note that the chosen metallicity distribution should be representative of the observed population to which this method is applied.

To account for measurement uncertainty in the observables, we adopt the following noise models:

\begin{enumerate}

\item H$\mathrm{\alpha}$ flux: Using the JADES DR3 emission-line catalog \citep{Eisenstein2023,Deugenio2024}, we derive uncertainties as a function of \Halpha\ flux. We include prism and grating measurements from GOODS-N/S with equal weight, restricting to sources with exposure times within 25\% of the median. Fluxes are logarithmically binned, and we compute the mean and 1$\sigma$ dispersion per bin.

\item Rest-frame NUV flux density: From the UNCOVER DR3 SUPER photometric catalog \citep{Bezanson2024,Suess2024,Weaver2024}, we compute NUV uncertainties using filters F090W, F115W, and F150W (rest-NUV at $z=4$). We restrict to measurements with depth weights within 5\% of the median. Data are logarithmically binned and weighted means and 1$\sigma$ spreads are computed. The weights follow $w=\int_{\lambda_1}^{\lambda_2} S(\lambda)\,d\lambda / (\lambda_2-\lambda_1)$, using sensitivity curves from \texttt{sedpy} \citep{Johnson2021_sedpy}.

\item Rest-frame FUV flux density: Same procedure as NUV, but using filters F070W and F090W (rest-FUV at $z=4$).

\item Balmer break strength: The uncertainty in Balmer break is largely independent of the break strength itself because it is defined as a ratio of continuum fluxes, causing the overall flux scaling to cancel and leaving the error dominated by continuum signal-to-noise rather than break size. Therefore, we simply adopt the mean and 1$\sigma$ dispersion of uncertainties from the Balmer breaks reported by \citet{Vikaeus2024}: $\mu_{\sigma,\text{break}} = 0.172$, $\sigma_{\sigma,\text{break}} = 0.094$.

\item U-V rest-frame color: Because U-V is a flux ratio (magnitude difference), its uncertainty is largely independent of the intrinsic color and is instead set by the photometric signal-to-noise in each band. From UNCOVER DR3, we compute U-V and its error for all entries with both values, yielding $\mu_{\sigma,\text{U-V}} \approx 2.1$ AB mag and $\sigma_{\sigma,\text{U-V}} \approx 0.8$ AB mag.

\end{enumerate}

\section{Simulation-Based Inference of Star Formation History Model Parameters}\label{sec:methods_sbi}

\begin{figure*}
    \centering
    \includegraphics[width=1\linewidth]{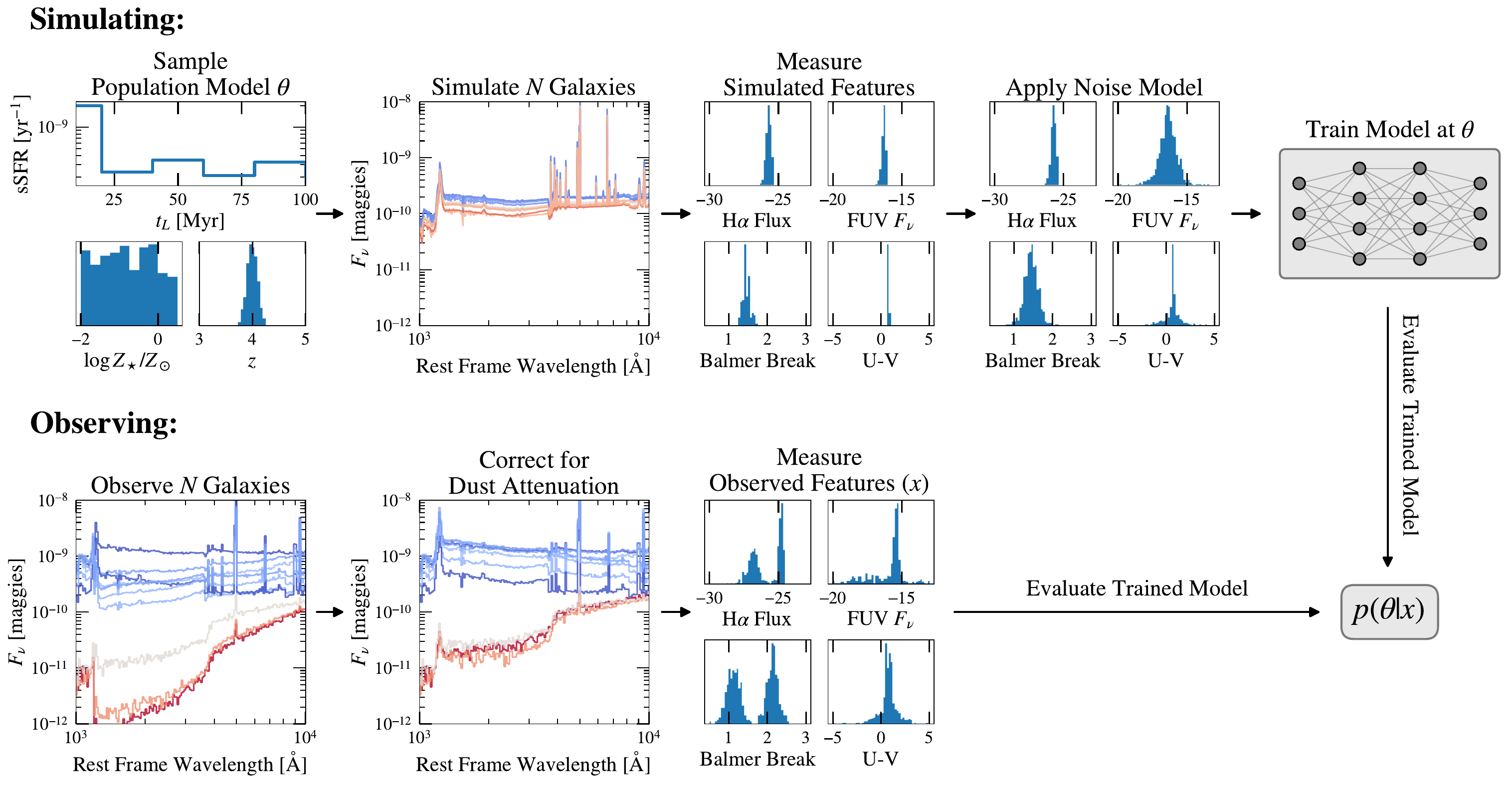}
    \caption{Workflow of the proposed method. A training set is generated by sampling parameters $\theta$ from a chosen population-level SFH model, including a distribution of stellar metallicities and redshifts in which to marginalize over. For a given sample of parameters, a population of $N$ SFHs is sampled from the SFH model, and the nuisance parameters are marginalized over. These parameters are passed to an SPS framework to model the observed spectrum of each object. Spectral features of interest are measured, with observational and counting noise applied. The SBI model is then trained to learn the mapping from these observable distributions to the posterior distribution of $\theta$. To apply this method to observed galaxy populations, $N$ observed spectra are corrected for dust attenuation, the same spectral features are measured, and the resulting distributions are passed to the trained model to infer posteriors on the selected population parameters.}
    \label{fig:sbi_pipeline}
\end{figure*}

We utilize SBI to recover population-level SFH parameters from galaxy populations. The inputs to the respective SBI model are the marginal distributions of spectral features and their associated uncertainties (including measurement and counting noise), while the outputs are the posteriors for the SFH parameters. Figure \ref{fig:sbi_pipeline} provides an overview of the workflow.

\subsection{Training Set Generation}

To train the SBI models, we generate synthetic galaxy populations using the SFH models described in Sections~\ref{sec:simple_model} and \ref{sec:psd_model}. For each population, SFHs are sampled according to the chosen model, marginalizing over stellar metallicity.

\subsubsection{Single-Frequency Model SFHs}

For the illustrative single-frequency SFH model, we independently sample the slope and oscillation parameters from uniform distributions: $\sigma \sim \mathcal{U}(0.1,1)$ dex, $\delta t \sim \mathcal{U}(0,100)$ Myr, and $\alpha \sim \mathcal{U}(0.0,0.3)$ in units of power-law index. The phase of the oscillation is sampled as $\phi \sim \mathcal{U}(0,2\pi)$. Each SFH is then propagated through the forward model to generate mock observables. This produces a diverse set of SFHs that explore the parameter space fully while remaining unbiased.

\subsubsection{Flex-PSD Model SFHs}

For the multi-frequency flex-PSD model, we construct SFHs by specifying a PSD via the flex-PSD model parameters and using the Wiener-Khinchin theorem to compute the auto-covariance function (ACF) at a given time $t$ over all frequencies $f$:

\begin{equation}\label{eq:ACF}
    \text{ACF}(t) =  \int_{-\infty}^{\infty} \text{PSD}(f)e^{2\pi if t} df.
\end{equation}

\noindent In this way, we treat star formation as a Gaussian stationary stochastic process. 

\cite{Wan2024} developed a framework to model stochastic SFHs in \prospector\ by modifying the default non-parametric SFH prior, known as the continuity prior \citep{Leja2019}, based on the work of \cite{Iyer2024}. This prior defines the probability distribution of the log ratio of the SFR in adjacent time bins,

\begin{equation}
    \Delta \log\text{SFR} = \log(\text{SFR}_i / \text{SFR}_{i+1}),
\end{equation}

\noindent and is designed to favor smooth SFHs by penalizing sharp transitions between bins. \cite{Wan2024} extended this approach to a stochastic prior, in which the log SFR ratios for each time bin are modeled as a multivariate normal distribution:

\begin{equation}\label{eq:stochastic_logsfr_ratios}
    \frac{\log\text{SFR}_i} {\log\text{SFR}_{i+1} } \sim \mathcal{N}(0,\text{ACF}_{i}(t)).
\end{equation}

\noindent With said framework, we adjust it such that the ACF from an arbitrary PSD can be numerically calculated via Equation \ref{eq:ACF} for use with the flex-PSD formalism.

The population-level parameters we infer include the five flex-PSD log-normalizations, as well as the slope parameters $\mu_\alpha$ and $\sigma_\alpha$, for a total of seven parameters describing the PSD. Each logPSD$_i$ is independently sampled from uniform priors, logPSD$_i \sim \mathcal{U}(-8,2)$ dex$^2$ Gyr, while the slope parameters follow $\mu\alpha \sim \mathcal{U}(-5,5)$ and $\sigma\alpha \sim \mathcal{U}(0,5)$ in power-law index units. SFHs are then generated by sampling from Equation \ref{eq:stochastic_logsfr_ratios}. This framework spans a broad range of possible SFHs, from smooth to highly bursty, allowing the SBI model to learn robustly across the parameter space. In practice, one may adopt priors restricted to physically allowable regions of PSD space. For example, excluding realizations that place large amounts of power only at very short timescales without the corresponding mass buildup on longer timescales, since such PSDs would be inconsistent with observed stellar mass growth.

\subsection{Calculating Observed Distributions of Spectral Features}\label{sec:methods_dists}

Synthetic SFHs are converted to observable distributions by forward-modeling through \texttt{FSPS} and \texttt{Prospector}, measuring spectral features of interest. The modeled spectra are not explicitly smoothed to account for instrumental resolution; instead, the noise model for each observable incorporates the full observational uncertainty. Each SBI model is trained on independent and uniformly sampled $1.1 \times 10^6$ populations ($10^6$ for training, $10^5$ for testing) across population-level parameter space, each with $N=500$ galaxy population members. This population size was selected to approximate the number of galaxy spectra that could realistically be obtained in a medium-to-large JWST program, while remaining large enough to suppress counting noise in the observable distributions. Feature distributions are computed with 50 bins per observable using \texttt{numpy.histogram}\footnote{We found that more complex density estimation methods, such as \texttt{scipy.stats.gaussian\_kde} \citep{2020SciPy}, were too computationally expensive to use for generating a large training set and bootstrapping density uncertainties (taking $\sim 1$ ms per evaluation, compared to $\sim 10 \ \mathrm{\mu s}$ for \texttt{np.histogram}).}
 \citep{numpy2020}. This binning choice represents a compromise between resolution and statistical noise for populations of size $N=500$. Measurement uncertainty is included via 500 bootstrap samples, drawing errors from truncated normal distributions $\in [0,\infty)$. We verified that the distribution uncertainty converges for this number of bootstrap realizations. For application to the joint distribution of spectral features, see Appendix \ref{app:joint}; for discussion on the choice of population size and its impact on the inference, see Appendix \ref{app:population_size}.

Counting noise is added by sampling bin counts $n$ from a Poisson distribution $n' \sim \mathcal{P}(\sqrt{n})$, and combined with measurement noise in quadrature. All distributions are then normalized to a unit integral. Distributions and uncertainties are flattened and concatenated into single input vectors before training the model. Simulating all $5.5 \times 10^8$ mock spectra and computing observed distributions requires $\sim 15,000$ CPU hours. 

\subsection{SBI Model Training and Posterior Evaluation}\label{sec:methods_sbitraining}

We infer the population-level parameters of our SFH models using SBI, which allows us to learn a neural density approximation of the posterior distributions of parameters directly from forward-modeled mock galaxy populations. In our approach, distributions of timescale-sensitive observables (including uncertainties) are passed through a normalizing flow, a flexible neural network that transforms a simple base distribution (e.g., a Gaussian) into a complex approximation of the true posterior. This allows us to capture correlations and non-Gaussian features in the posterior that would be difficult to model analytically.

 The setup of the SBI models in this work generally follows that of \citet{Hahn2022} and \citet{Wang2023:sbipp}. The SBI models are implemented using Masked Auto-regressive Flows \citep[MAFs;][]{Papamakarios2018} via the \texttt{sbi} Python toolkit \citep{TejeroCantero2020}, and trained using Sequential Neural Posterior Estimation \citep[SNPE;][]{Greenberg2019}. SNPE works iteratively: at each step, a set of parameter values is sampled from the prior, synthetic observables are generated via the forward model, and the network is trained to map these observables to their corresponding parameters. Subsequent rounds focus on regions of parameter space where the posterior probability is highest, improving sample efficiency and posterior accuracy. For our models, the MAFs consist of 15 blocks, each with two hidden layers of 500 units. Posterior evaluations are performed by sampling 1,000 times from the trained flow unless stated otherwise.\footnote{Training time per model is $\sim 3-4$ hours on an NVIDIA A100 GPU.}

We quantify posterior median accuracy using the normalized median absolute deviation (NMAD):

\begin{equation}
    \sigma_\text{NMAD} = 1.48 \times \text{Median}\big(|\tilde{\theta}_\text{pred} - \theta_\text{true}|\big),
\end{equation}

\noindent where $\tilde{\theta}_\text{pred}$ is the posterior median for a given hyperparameter. NMAD provides a robust measure of the typical error between the inferred and true parameters, even in the presence of non-Gaussian posteriors.

\subsection{Mock Populations for Testing}\label{sec:test_populations}
 
To assess the robustness of our approach, we apply the method to mock galaxy populations whose SFHs differ from those assumed in training, including populations motivated by cosmological simulations as well as synthetic populations designed to capture realistic patterns of SFHs. This enables us to quantify the effects of model mismatch under a range of physically plausible conditions. In the sections below, we describe the feedback prescriptions and variability characteristics that motivate each population, and outline how we generate the corresponding observables to which our methodology is applied. Together, these populations span a wide range of SFH variability, from highly bursty to smooth, enabling a stringent test of model robustness under realistic model mismatch.

\subsubsection{High-$z$ FIRE-2-like Population}\label{sec:fire}

The FIRE-2 simulations \citep{Hopkins2014:fire,Hopkins2018:fire2} are a set of cosmological zoom-in simulations that model galaxy formation, including gas dynamics, radiative cooling, and the stochastic formation of star particles \citep{Hopkins2014:gizmo,Ferland2017}. Stellar feedback in FIRE-2 includes radiation pressure, photoionization, stellar winds, and kinetic energy injection from core-collapse and Type Ia supernovae, but feedback from supermassive black holes is not included \citep{Hopkins2014:fire,Hopkins2018:fire2}. The zoom-in approach allows FIRE-2 to resolve the multiphase interstellar medium (ISM) at smaller spatial scales than simulations like Illustris. 

We note that the SFHs in FIRE-2 are constructed from discrete star particles, and the finite particle mass can cause the SFH to appear burstier than an otherwise smooth underlying SFH, particularly when not binned to longer time intervals. Additionally, the galaxies in the high-$z$ FIRE-2 suite are generally lower-mass than those we model in this work. For these reasons, the FIRE-2-like SFHs we use as a test population in this paper can appear artificially more bursty than the original FIRE-2 simulations when evaluated at fixed stellar mass. However, this does not impact our goals, as we are using these populations purely to test our framework by comparing two extreme cases: highly bursty versus smooth SFHs.

We consider a sample of 506 galaxies from the high-$z$ FIRE-2 suite, specifically the \texttt{z5m11X} and \texttt{z5m12X} series \citep{Ma2018, Ma2019}, with SFHs binned to 1 Myr and stopping at $z=5$. These SFHs will be presented in an upcoming paper by Gonzales et al. (in prep). To ensure repeatability and uniformity in testing our method, we first measure the PSD of these SFHs, average it across the population, and then use it to generate a new population of $N=500$ SFHs at fixed stellar mass ($M_\star = 10^{10} \ \mathrm{M_\odot}$) to create a ``FIRE-2-like" testing population. From the original FIRE-2 SFHs, we calculate the mean and dispersion of the best-fit power-law slopes over $t_L \leq 500$ Myr, finding $\mu_\alpha \approx 0.1$ and $\sigma_\alpha \approx 1.1$. We then simulate observed spectra and compute spectral feature distributions for the PSD-resampled population following the methodology in Section \ref{sec:methods_sps} to avoid differing systematics in SPS modeling. Hereafter, the ``FIRE-2-like" testing population is representative of the low-mass, high-redshift galaxies seen in the FIRE-2 simulations, and is an idealized representation of a highly bursty galaxy population.

\subsubsection{Low-$z$ Illustris-like Population}\label{sec:illustris}

Illustris \citep{Genel2014,Vogelsberger2014b} is a large-scale cosmological hydrodynamical simulation that models gas physics including primordial and metal-line cooling, gas recycling, and stochastic star formation \citep{Springel2003,Vogelsberger2013}. Stellar feedback is implemented via radiation pressure, local photoionization, heating, and winds, with supernovae and supermassive black hole kinetic feedback also included \citep{Springel2005:bh_feedback,Sijacki2007,Pillepich2018b}. Illustris has a spatial resolution of $\sim 1$ kpc, which does not resolve the multiphase ISM at the scale of individual giant molecular clouds. Its model parameters are tuned to reproduce observed galaxy scaling relations, including the cosmic star formation rate density, the galaxy stellar mass function, and the stellar mass-halo relation \citep{Schmidt1959,Kennicutt1989}. Compared to the FIRE-2 sample, Illustris galaxies are generally higher-mass, and their SFHs reflect larger-scale behavior at low redshifts.

The temporal resolution of Illustris SFHs is non-uniform, and can range from $<10$ Myr to $\sim 100$ Myr. PSD estimation via Welch's method requires a uniformly sampled time series; therefore, SFHs must be binned to the coarsest effective time resolution, suppressing power on short timescales. As a result, SFH variability on $\lesssim 100$ Myr timescales cannot be robustly recovered, preventing a direct comparison between Illustris and FIRE-2 PSDs.

\citet{Iyer2020} found that the Illustris PSD approximately follows a power-law with slope $\beta \sim 2$ on 100 Myr-1 Gyr timescales. For simplicity in this validation exercise, we adopt this power-law slope across all timescales to create a representative ``Illustris-like'' population for method validation, while noting that the true PSD likely exhibits a break at short timescales. Using this PSD, we assume $\mu_\alpha = 0$ and $\sigma_\alpha = 0$, simulate a population of $N=500$ SFHs, generate corresponding mock spectra, and compute distributions of spectral features following the methodology in Section \ref{sec:methods_sps}. Hereafter, the ``Illustris-like'' testing population is representative of the high-mass, low-redshift galaxies seen in the Illustris simulations, and is an idealized representation of a non-bursty galaxy population.

\subsection{PSDs of Realistic Galaxy Populations}\label{sec:methods_realistic}

To evaluate model performance on galaxy populations with plausible levels of short- and long-timescale variability, we simulate 10$^3$ mock galaxy populations, each with a population SFH model intended to approximate the shapes of PSDs found in simulations and theory. We refer to these populations as ``Realistic" mock galaxy populations. ``Realistic" in this context refers to PSDs that roughly resemble both the PSDs measured in cosmological simulations \citep{Iyer2020}, and those that are theoretically predicted from analytical models of the processes that regulate star formation \citep{Tachella2020}. In all such cases, the PSD rises toward long timescales as an approximate broken power-law. The slope of this power-law, $\beta$, on $>100$ Myr timescales in simulations is $\beta \sim 2$ on timescales of $\gtrsim 100$ Myr \citep{Iyer2020}. To emulate the shapes of these PSDs in the flex-PSD formalism, we select flex-PSD model parameters that correspond to a broken power-law, with normalizations spanning the dynamic range of the SBI training set. Crucially, we model populations that have high-amplitude bursts of star formation (total power of the PSD is maximized) and have short-timescale fluctuations (shallow PSD). We additionally model the slope and slope dispersion of the recent SFH by independently and uniformly sampling them from the training set hyperparameter distribution. We calculate the observable distributions from these mock populations and evaluate the SBI model trained on the flex-PSD and sloped SFH parameters. 

\section{Results}\label{sec:Results}

In this section, we first examine how the distributions of observables respond to changes in the population SFH models introduced in Section \ref{sec:SFHModels}, and then evaluate the performance of the SBI models in recovering the underlying population-level SFH parameters using the methodology described in Sections \ref{sec:methods_population_modeling} and \ref{sec:methods_sbi}.

\subsection{Distribution of SFR Timescale Indicators Constrains Burstiness}

\begin{figure*}
    \centering
    \includegraphics[width=1\linewidth]{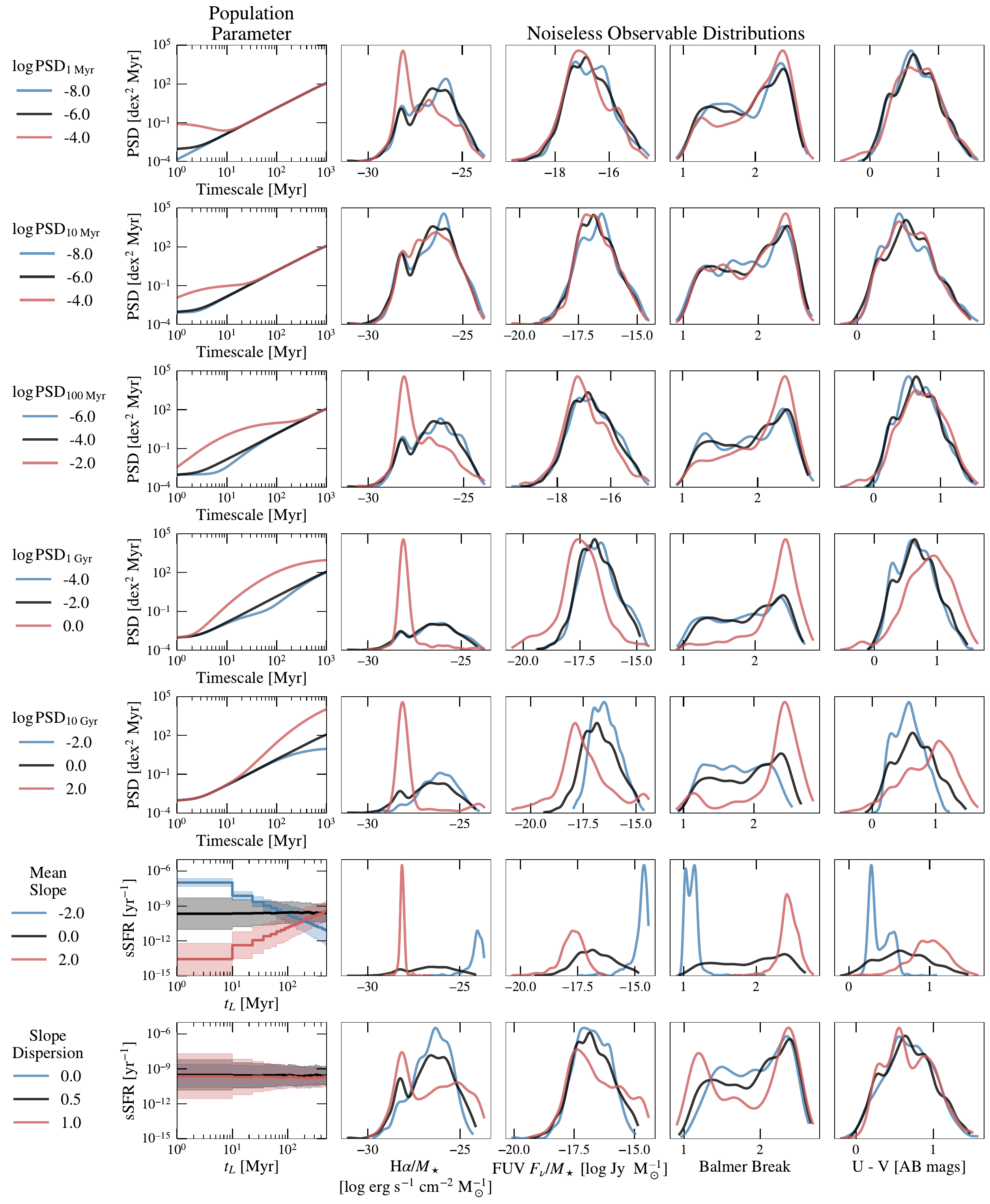}
    \caption{Effect of flex-PSD and SFH slope parameters on observable distributions. Each row displays the change in the PSD/SFHs and the observable distributions as we decrease (blue) or increase (red) the parameter by some amount. The black curve is the same population model in all panels. }
    \label{fig:flexpsd_model_mega}
\end{figure*}

The left panel of Figure \ref{fig:simple_model} shows three representative single-frequency SFH model realizations. By randomly sampling over the phase for $N=500$ galaxies, we simulate an observed population. The features considered here include the \Halpha\ flux (normalized by the total mass formed), the Balmer break strength, and the \Halpha/UV emission ratio. 

The \Halpha/UV\ distribution clearly distinguishes models with different burst amplitudes ($\sigma$). However, for two models with the same burst amplitude but differing timescales ($\delta t$), \Halpha/UV\ alone is insufficient to separate the populations. In other words, \Halpha/UV\ traces the intensity of recent bursts, but not their duration. By contrast, the marginal distribution of normalized \Halpha\ flux (top-right panel of Figure \ref{fig:simple_model}) differentiates both burst strength and timescale. The Balmer break, which probes $\sim100$ Myr timescales, shows little variation between these realizations, providing minimal sensitivity to short-timescale fluctuations, as it only becomes a significant tracer during periods of extended quiescence.

These results are consistent with \citet{Wang2025:sfh}, and demonstrate the need to combine observables probing a wide range of timescales. Short-timescale indicators (e.g., \Halpha\ and the UV continuum) and long-timescale features (e.g., the Balmer break, optical colors) together allow inference of SFR variations across the population. This approach is effective for both bursty populations dominated by rapid fluctuations and smoother populations where long-term trends prevail.

The flex-PSD model, illustrated in Figure \ref{fig:flexpsd_model}, produces a wide diversity of SFHs. This diversity arises because the SFHs are stochastic, and the flex-PSD model parameters can generate a wide array of different (although not necessarily physical) PSDs. The resulting SFHs range from highly bursty to relatively smooth, as displayed in the top right panel of Figure \ref{fig:flexpsd_model}. Importantly, this diversity is directly reflected in the distributions of observables: the marginal distributions of \Halpha, UV flux, Balmer break strength, and rest-frame U-V color are noticeably distinct between bursty and smooth populations. This demonstrates that the flex-PSD formalism captures a broad range of SFH behaviors and produces corresponding observable signatures, making it well-suited for the population-level inference methodology presented in this work. We note, however, that the flex-PSD model exhibits some inherent degeneracies between its parameters. For example, when a given Gaussian component has a large normalization, it can wash out the signatures of surrounding components, producing a similar PSD if those components have significantly lower normalizations (see second row, first column of Figure \ref{fig:flexpsd_model_mega}). While these degeneracies cannot be resolved from the distribution of observables, this is not a concern. For our purposes, the overall shape of the PSD matters most, rather than the precise values of the individual parameters.

In addition to the illustrative bursty and smooth models shown in Figure \ref{fig:flexpsd_model}, the shapes of the observable distributions are highly sensitive to the chosen PSD and SFH slope parameters, as illustrated in Figure \ref{fig:flexpsd_model_mega}. Each observable responds most strongly to PSD variations at the timescales it probes: the Balmer break to $\gtrsim100$ Myr, and rest-frame U-V color to $\gtrsim1$ Gyr. Note that higher PSD normalizations correspond to larger amplitude fluctuations, meaning that a larger fraction of the population is likely to be observed in low-SFR phases. The slope parameters, in particular, are tested here as extreme cases to illustrate their effects; we do not necessarily expect real populations to have such steeply rising/falling SFHs. They dramatically skew the observable distributions, creating asymmetry that is not a result of the burstiness of the SFHs. This emphasizes that a population's SFH slope must be modeled separately, as changes in it can mimic or obscure signatures of short-term fluctuations in star formation.

\subsection{Single-Frequency SBI Model Performance}\label{sec:simple_results}

\begin{figure*}
    \centering
    \includegraphics[width=1\linewidth]{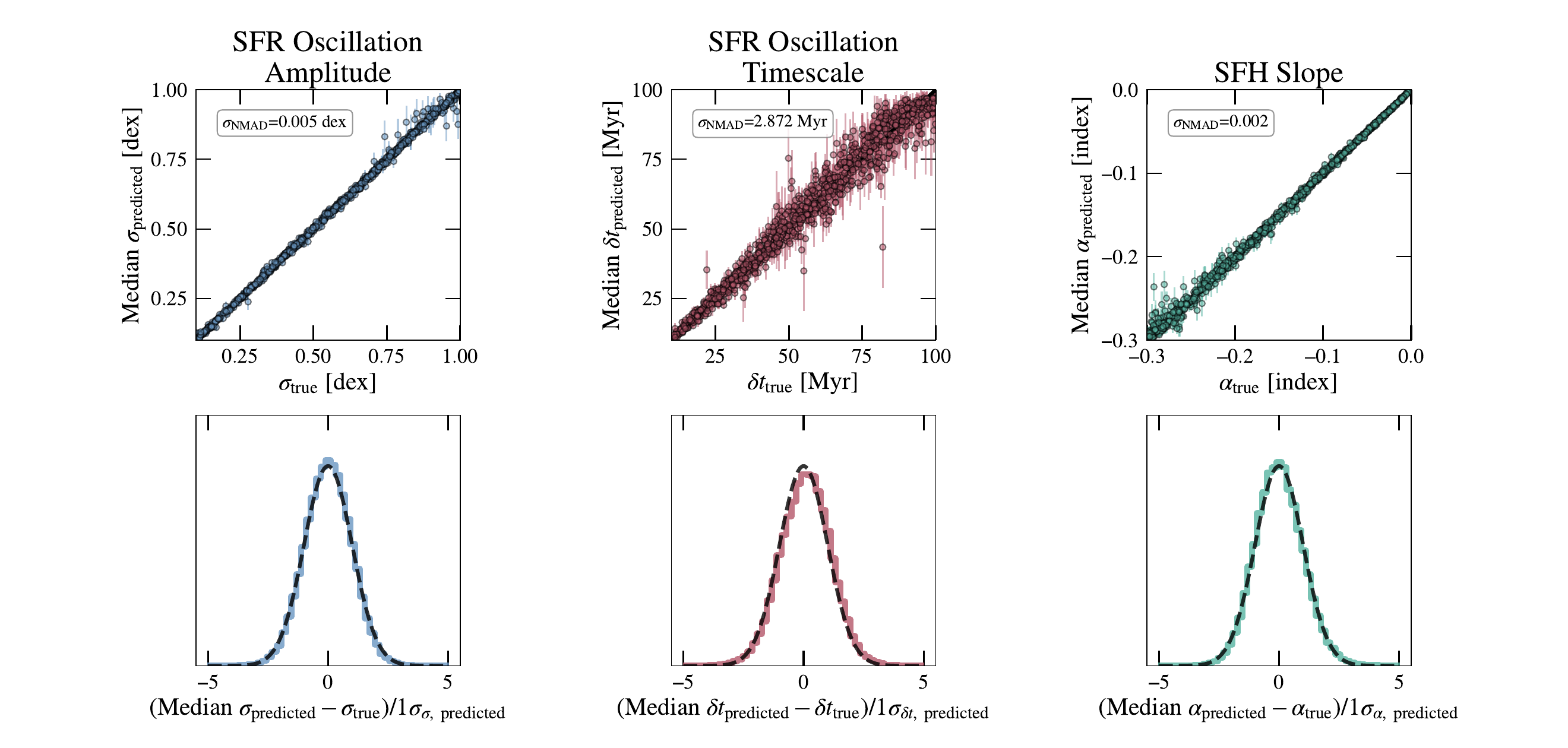}
    \caption{SBI model residuals for the single-frequency SFH model. Upper Row: Posterior median $\pm 1\sigma$  versus the true value of the hyperparameter. From left to right, $\sigma$, $\delta t$, and $\alpha$. A one-to-one line is displayed as a black dashed line for reference. $\sigma_\text{NMAD}$, the normalized median absolute deviation, is shown in the top left corner. Lower Row: Distribution of the standardized residuals for each simple SFH hyperparameter. A unit-Gaussian is displayed as a black dashed line for reference.}
    \label{fig:simple_model_results}
\end{figure*}

In this section, we present the results of an SBI trained on the single-frequency population SFH model, as introduced in Section \ref{sec:simple_model}. Figure \ref{fig:simple_model_results} presents the recovery of single-frequency SFH parameters for $10^5$ mock testing populations. To generate these results, we create mock populations and their spectra, measure timescale-sensitive features, apply realistic measurement noise models, evaluate histograms of the observables, and simulate counting noise. We then evaluate the SBI model on the density values of these histograms, which returns the posteriors for the three single-frequency model parameters ($\sigma$, $\delta t$, and $\alpha$). The top row shows the predicted posterior medians versus the true hyperparameter values, while the bottom row displays the distributions of standardized residuals across the test set. 

These results demonstrate that, in the single-frequency regime, the proposed methodology accurately recovers population-level SFH parameters from the marginal distributions of observables. Burst amplitudes are recovered with precision better than $0.1$ dex, overall SFH slopes within $<0.1$, and burst timescales within $\sim 3$ Myr. Furthermore, the standardized residuals are approximately Gaussian, indicating that the posterior uncertainties are well-calibrated. This demonstrates that the methodology performs exceptionally well in the simple-frequency SFH regime, accurately recovering population-level SFH parameters and validating the approach as a proof-of-concept.

\subsection{PSDs Recovered from Cosmological Simulations}\label{sec:results_psd}

\begin{figure*}
    \centering
    \includegraphics[width=1\linewidth]{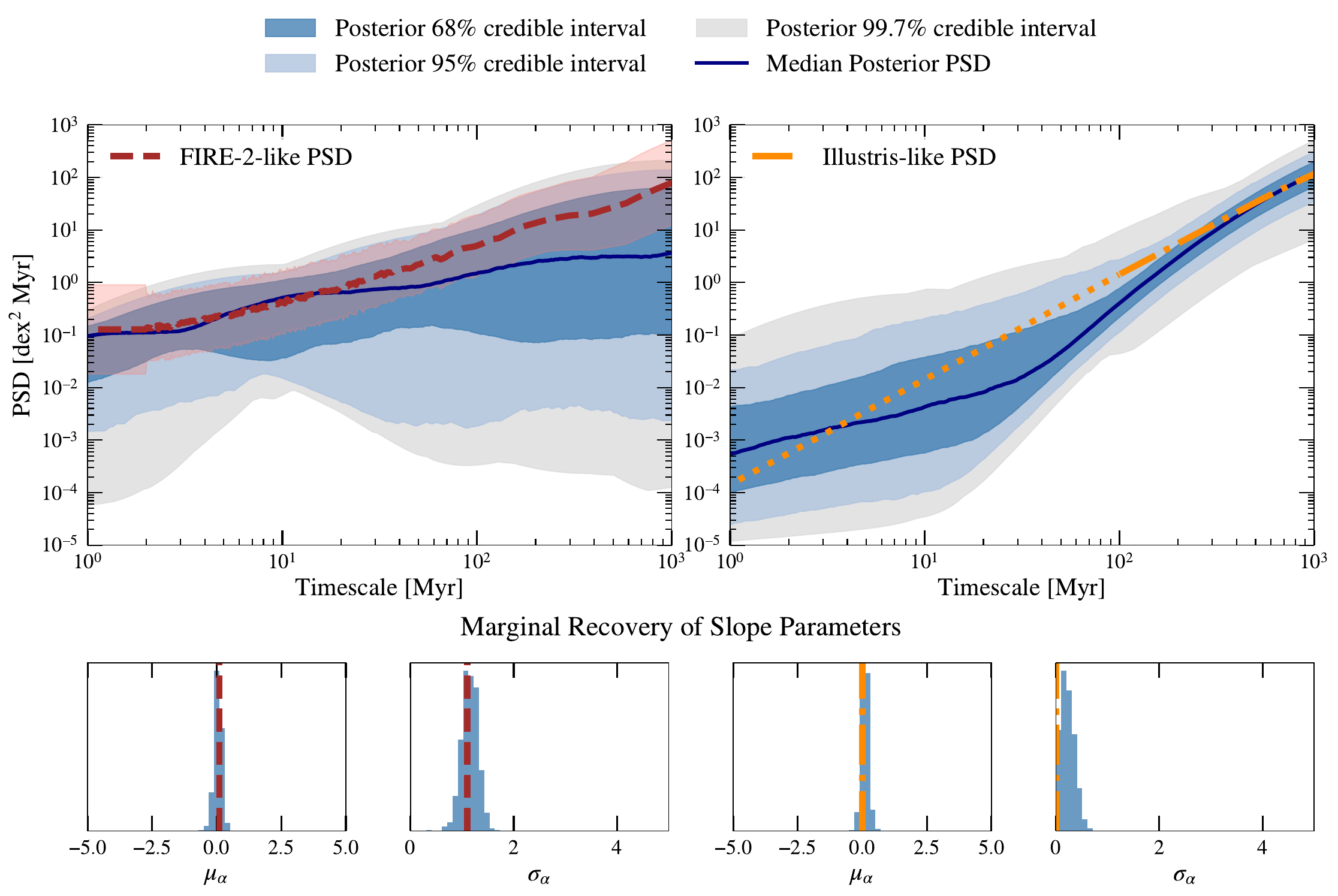}
    \caption{Recovery of the FIRE-2-like and Illustris-like population PSDs and slope parameters. Top panels: Recovery of the PSDs for FIRE-2-like (left) and Illustris-like (right) populations. The FIRE-2-like PSD is the red dashed line, and the dispersion of the calculated FIRE-2-like PSD is shaded in red. The Illustris-like PSD is shown with an orange dot-dashed line (the dotted region indicates that the true power of the Illustris project SFR fluctuations cannot be calculated due to the limited time resolution). The posterior median PSD is shown as a dark blue solid line, and the posterior 68\%, 95\%, and 99.7\% confidence intervals are shaded in blue, light blue, and gray, respectively. Lower panels: Marginal posterior probability distributions for the mean and dispersion of SFH slopes for FIRE-2-like and Illustris-like populations. The true value of the parameters is indicated with a red dashed line for FIRE-2-like, and an orange dot-dash line for Illustris-like. For a visualization of the marginal posteriors of the flex-PSD parameters recovered for the FIRE-2-like PSD, see Figure \ref{fig:population_size} in Appendix \ref{app:population_size} for a sample size of $N=500$.}
    \label{fig:fire_illustris_results}
\end{figure*}

In this section, we present the results of an SBI model trained on the flex-PSD population SFH model with sloped SFHs, as introduced in Section \ref{sec:psd_model}. To generate these results, we create mock populations whose SFHs resemble FIRE-2 and Illustris galaxies, simulate their spectra, measure timescale-sensitive features, apply realistic measurement noise models, evaluate histograms of the observables, and simulate counting noise. The trained SBI model is then evaluated on these histogram densities to obtain the posterior distributions of the flex-PSD parameters (five logPSD$_i$ normalizations) as well as the mean slope and slope dispersion of the populations' SFHs ($\mu_\alpha$ and $\sigma_\alpha$). We project the resulting posteriors of the flex-PSD parameters into Fourier space, as shown in the top panels of Figure \ref{fig:fire_illustris_results}. The marginal posterior distributions for the SFH slope parameters are displayed in the lower panels.

For the bursty FIRE-2-like population, the SFR fluctuation power is recovered accurately: the input PSD (red dashed line) lies within the 68\% posterior confidence interval over timescales from 1 Myr to 10 Gyr. A bias is observed at $\gtrsim 100$ Myr timescales, where the posterior median underestimates the true FIRE-2-like PSD by $\sim 1$ dex; while this is small compared to the $1\sigma$=1.5 dex uncertainties, it is significant. We discuss potential origins of this bias in Section \ref{sec:discussion_outshining}.

The Illustris-like population PSD is also well recovered. The true PSD falls within the 68\% posterior interval for 1-20 Myr and 200 Myr-1 Gyr, and within the 95\% interval for intermediate 20-200 Myr timescales. Compared to FIRE-2-like, the posterior widths at long timescales ($>100$ Myr) are much narrower, with the central 68\% confidence interval spanning only $\sim 0.5$ dex, versus $\sim 2.8$ dex for FIRE-2-like. This indicates that the Illustris-like PSD is recovered with higher confidence than the FIRE-2-like population. A potential explanation for this difference is discussed further in Section \ref{sec:discussion_outshining}. Crucially, each fit is sufficiently confident to rule out the other model, suggesting that a medium-to-large JWST/NIRSpec survey ($\sim$500 spectra) of a uniform, complete, and closely-defined sample of comparable galaxies should be able to distinguish cleanly between these different numerical models of star formation. See Appendix \ref{app:population_size} for results of how recovery accuracy varies as a function of sample size.

In both populations, the mean and dispersion of the SFH slope are precisely and accurately recovered within the $1\sigma$ posterior intervals. This accuracy arises from the strong correlation between these slope parameters and the skewness of the observable distributions, as illustrated in the lower rows of Figure \ref{fig:flexpsd_model_mega}.

\subsection{Recovery of PSDs in Realistic Mock Galaxy Populations}\label{sec:results_outshining}

\begin{figure*}[htbp]
\centering
\gridline{
    \fig{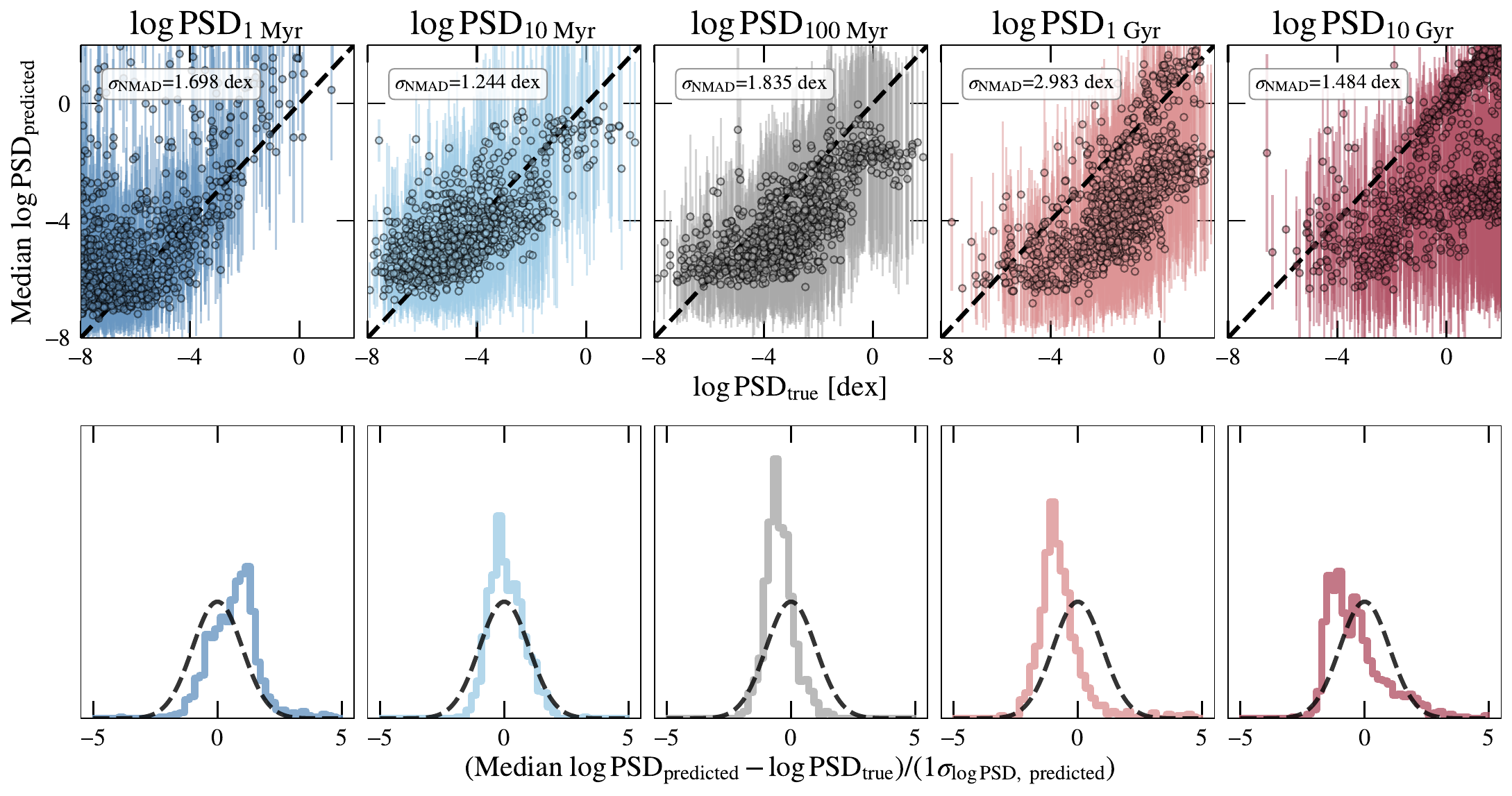}
    {1.0\textwidth}{(a) SBI model residuals for the flex-PSD parameters.}
}
\gridline{
    \fig{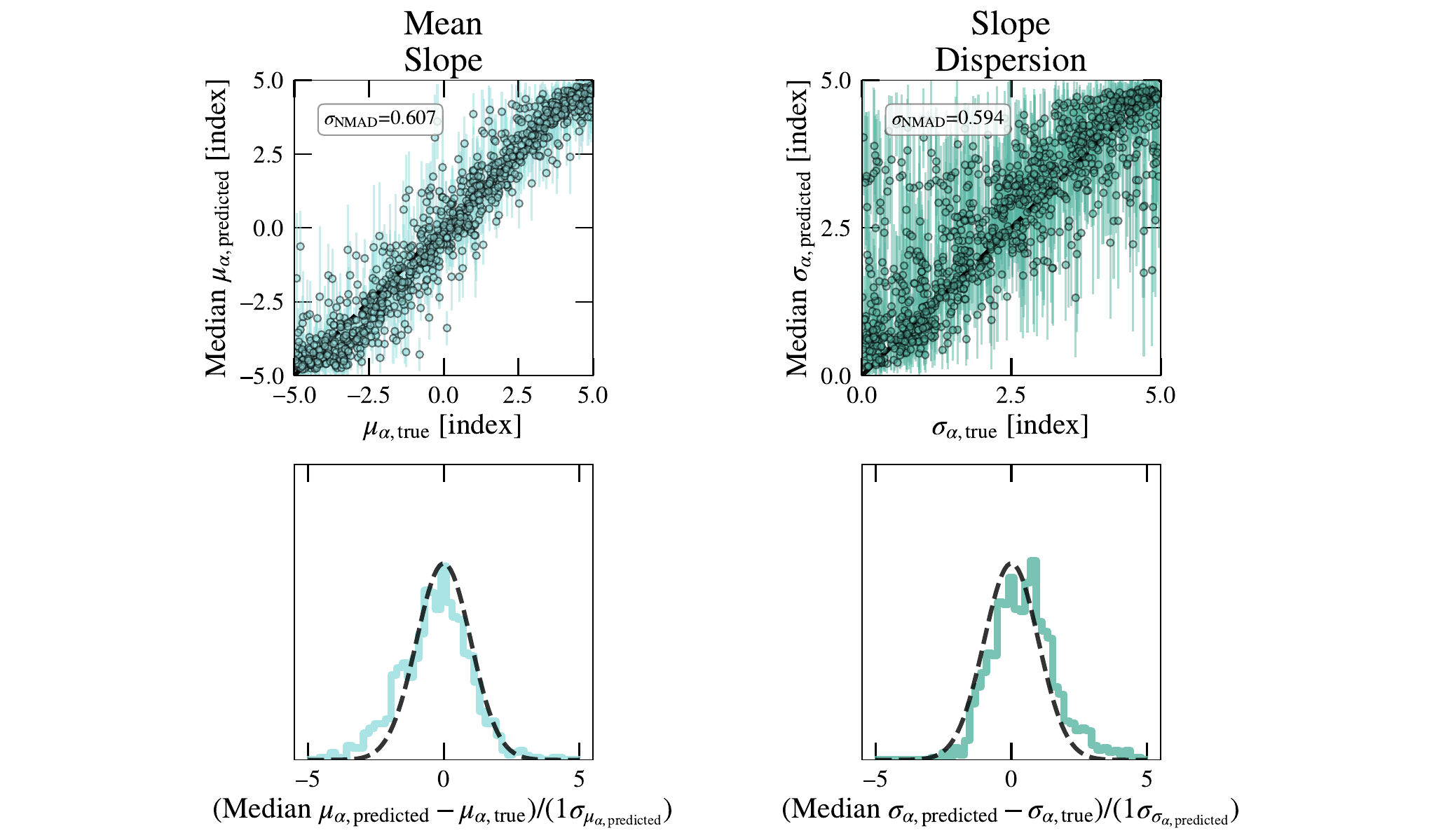}{1.0\textwidth}
    {(b) SBI model residuals for the population SFH slope parameters.}
}
\caption{SBI model residuals for a ``realistic'' flex-PSD population model parameters. In both panels, the upper row displays the median posterior $\pm 1\sigma$ versus the true value of the hyperparameter; a one-to-one line is displayed as a black dashed line for reference. The lower row displays the standardized residuals for each hyperparameter. A unit-Gaussian is displayed as a black dashed line for reference.}
\label{fig:flexpsd_model_realistic_results}
\end{figure*}

\begin{figure}
    \centering
    \includegraphics[width=1\linewidth]{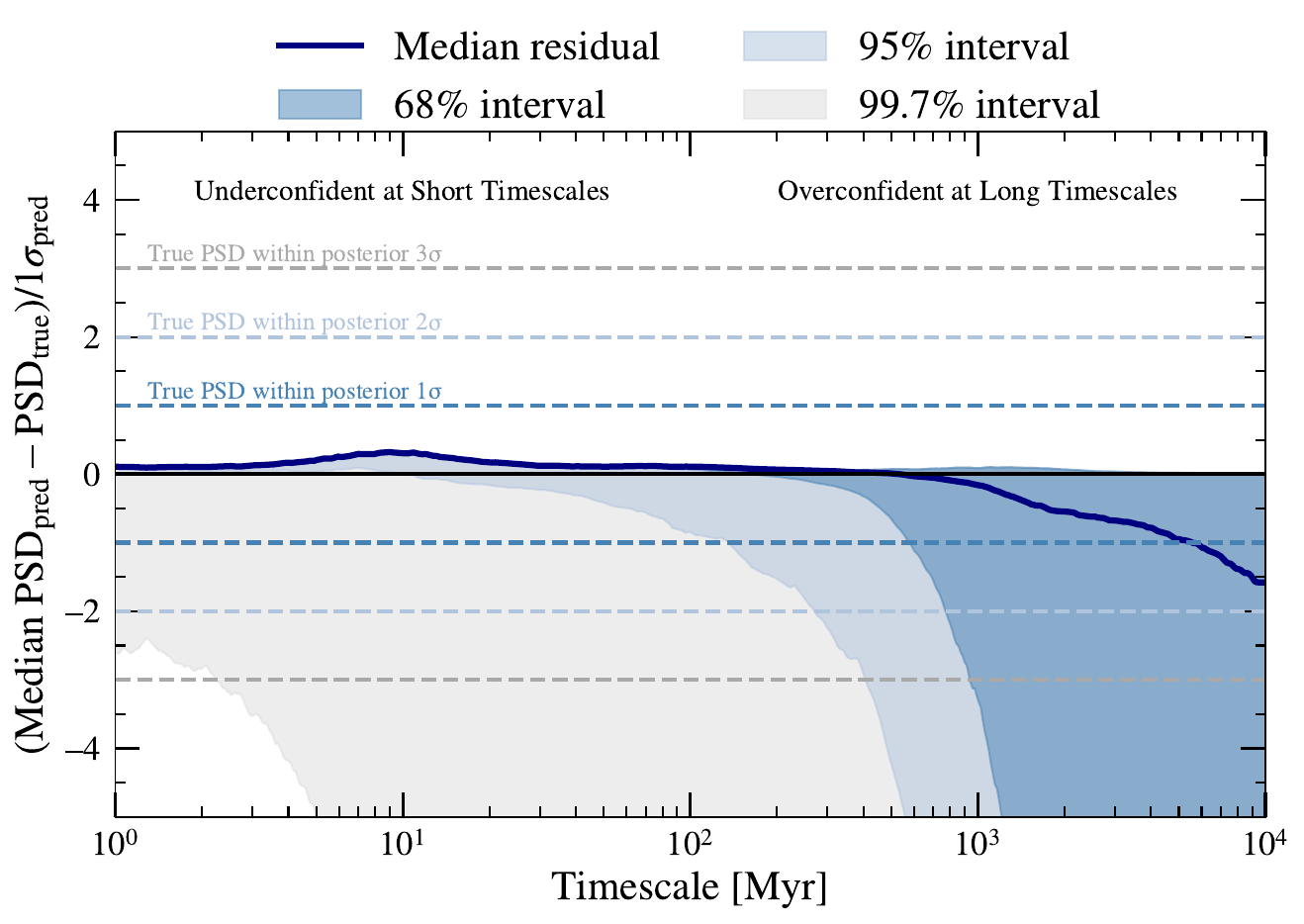}
    \caption{Residuals for the inferred PSD from our ``realistic" flex-PSD population. We display the median residual (dark blue line) as a function of timescale for mock populations' PSDs. The shaded regions correspond to 68\%, 95\%, and 99.7\% intervals for the standardized residuals (blue, light blue, and gray shades). The blue, light blue, and gray dashed lines bracket the area that indicates that the true PSD is within $1\sigma$, $2\sigma$, and $3\sigma$  dispersions of the posterior PSD, respectively.}
    \label{fig:flexpsd_model_realistic_results_psd}
\end{figure}

We further evaluate the SBI model trained on the flex-PSD formalism on the ``realistic" galaxy populations discussed in Section \ref{sec:methods_realistic}. It is important to note that these populations do not follow the prior distribution of the training data for this population. The SBI model is trained on flex-PSD parameters that are independently and uniformly sampled over a broad dynamic range, and therefore include PSD shapes that may not resemble physically motivated SFHs. This ensures that the model is not biased toward said populations, a consequence we discuss further in Appendix \ref{sec:app_exreg}. For discussion on the model recovery of mock populations that follow the training data, see Appendix \ref{sec:app_testset}.

The results of recovering the parameter values for the ``realistic'' mock populations are displayed in Figure~\ref{fig:flexpsd_model_realistic_results}. The SFH slope parameters are the most accurately constrained of all the parameters. For both parameters, $\sigma_\text{NMAD} <1$, indicating a rather low scatter in residuals. The distribution of their standardized residuals is approximately normal around zero, with only a slight skew toward under-predicting the mean slope and overestimating the scatter, implying that the model is well-calibrated for these parameters. This is expected, as the slope parameters have the largest impact on the observable distributions that are fed into the model and are the primary features on which the SBI model is trained. 

In contrast, the PSD parameters are recovered less reliably. There is a clear bias to slightly over-predict the power of 1 Myr fluctuations and to under-predict the power at 1-10 Gyr timescales by roughly $1\sigma$-$2\sigma$. While the posterior medians are often within a few $\sigma$ of the true value, the distribution of standardized residuals for this ``realistic'' population is not normally distributed, indicating poorer calibration. The $\sigma_\text{NMAD}$ values for the PSD parameters range from 1.2-3.0 dex, reflecting high scatter in the residuals. 

Because these hyperparameter residuals are difficult to interpret directly, it is more informative to examine the residuals of the inferred PSD. Figure \ref{fig:flexpsd_model_realistic_results_psd} shows the distribution of the standardized residuals in the PSD as a function of timescale. The shaded regions display the 68\%, 95\%, and 99.7\% confidence intervals for the distribution of PSD residuals (i.e., the distribution of $\chi = (\text{Median }\mathrm{PSD}\text{pred} - \mathrm{PSD}\text{true}) / 1\sigma_\text{pred}$ as a function of timescale for all mock ``realistic'' populations). The black solid line indicates perfect recovery of the PSD at a given timescale. For a well-calibrated model, we expect the gray shaded regions to approximately follow the 1, 2, and 3$\sigma$ values of $\chi$.

In practice, the central 68\% interval of the $\chi$ distribution is near zero at timescales $\lesssim 100$ Myr, indicating that the model is under-confident at short timescales. The 95\% interval is broader but also under-confident, while the 99.7\% interval is very wide and appears over-confident at nearly all timescales above 10 Myr. At timescales $\gtrsim 100$ Myr, all intervals (68\%, 95\%, and 99.7\%) are wider than expected, indicating over-confidence in the recovery of long-timescale power. Notably, the distributions rarely occupy positive $\chi$ values, meaning that $(\text{Median }\mathrm{PSD}\text{pred} - \mathrm{PSD}\text{true}) / 1\sigma_\text{pred}$ is almost always negative. This behavior arises from the non-Gaussian structure of the posteriors, which are often skewed toward low power and act more as upper limits on the maximum PSD power at a given timescale. We attribute this to the inherent degeneracy of the flex-PSD model and the fact that the flex-PSD parameters are uniformly and independently sampled from the training prior, which allows for any large values for any individual flex-PSD parameter to wash out signals from the other Gaussian components (see Figure \ref{fig:flexpsd_model_mega}). In summary, taking the median posterior PSD as the inferred PSD leads to a likely under-prediction of power at long timescales, while the power at short timescales remains well-constrained. Another potential explanation for the poor calibration at long timescales is that we have reached a fundamental observational limit for understanding the $>100$ Myr timescale variations in galaxy SFHs. We explore this further in Section \ref{sec:discussion_outshining}.

\section{Discussion}\label{sec:discussion}

Here, we place our results into the broader context of modeling and inferring galaxy SFHs. We begin by examining the importance of explicitly including the population-averaged SFH slope to accurately recover both short- and long-timescale power of SFR fluctuations. We then consider how highly bursty populations can obscure long-timescale features in the PSD. Finally, we conclude by outlining practical considerations for applying this methodology to observed spectroscopic samples, including criteria for suitable populations and the potential uses of inferred PSDs from observed galaxy populations.

\subsection{Modeling the Recent SFH Slope}\label{sec:discussion_slope}

\begin{figure*}[htbp]
\centering
\gridline{
    \fig{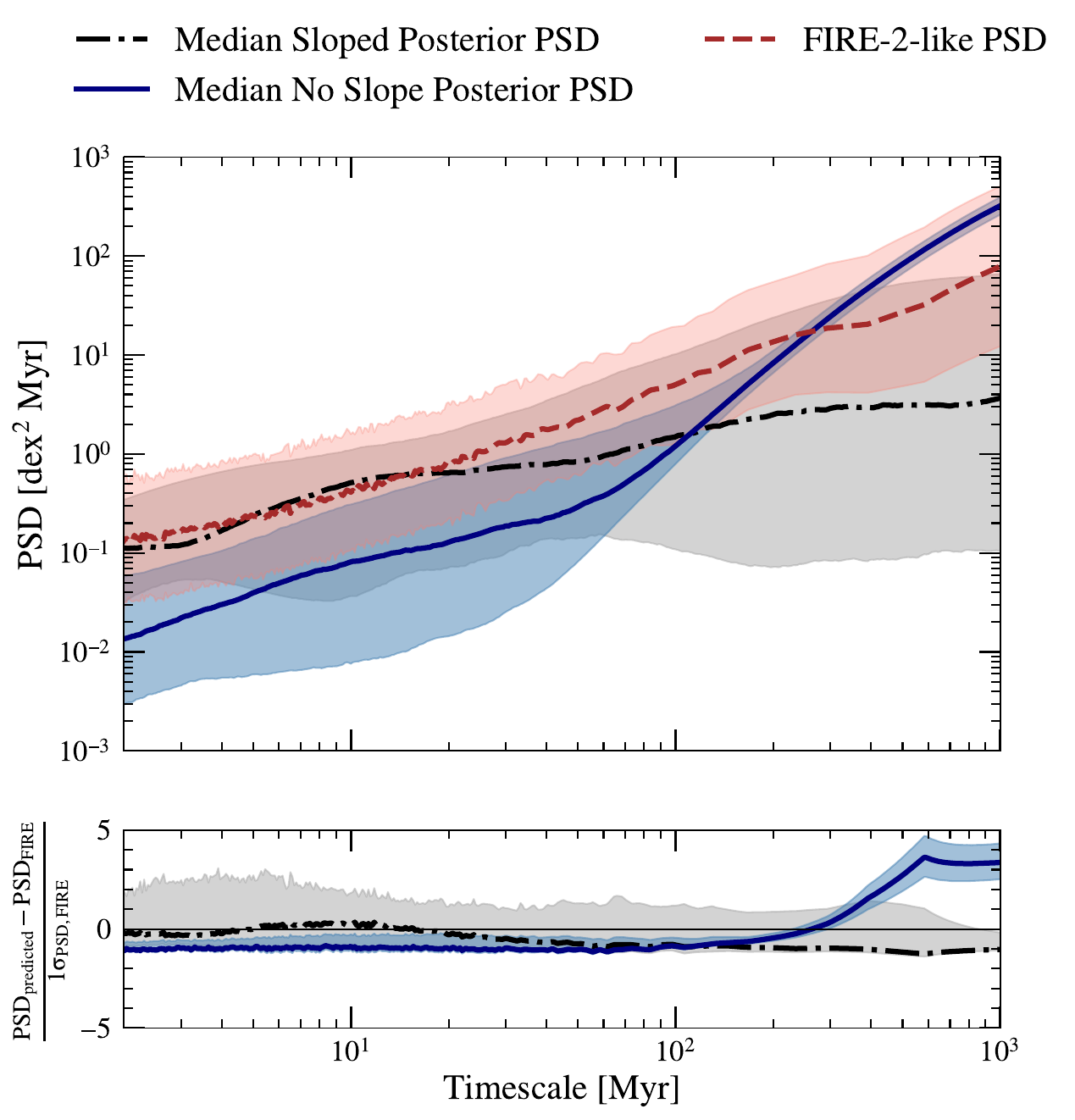}{0.49\textwidth}{(a)}
    \fig{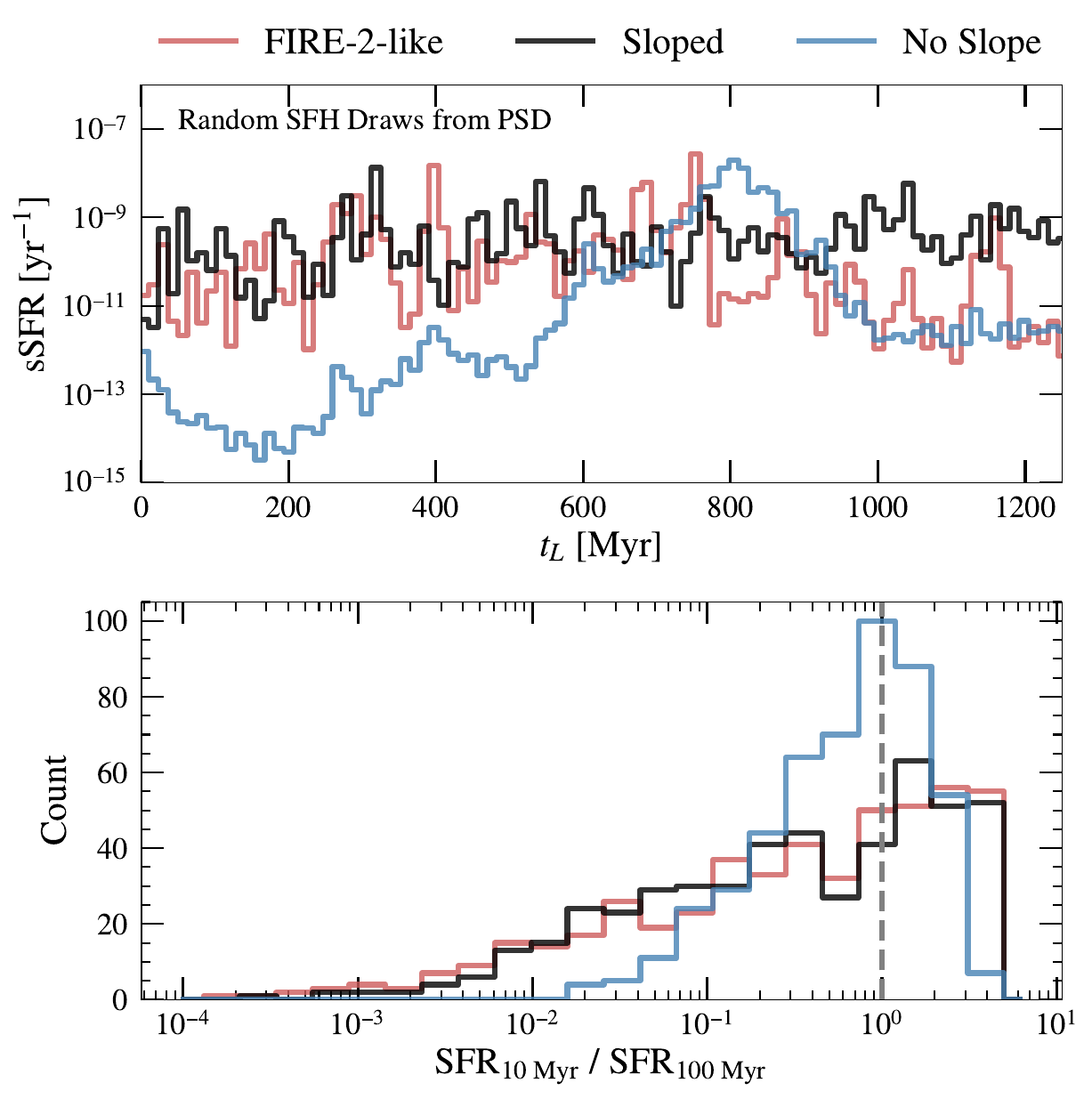}{0.49\textwidth}{(b)}
}
\caption{Recovery of the FIRE-2-like PSD with the SBI model trained without sloped SFHs. 
Left panel: The upper panel displays the recovered PSD from an SBI model trained without sloped SFHs (blue solid) and with sloped SFHs (black dot-dashed). The shaded areas indicate the posterior 68\% confidence interval. The true FIRE-2-like population PSD is shown as a red dashed line, and the red shaded region corresponds to the dispersion of the populations' PSDs. The lower panel displays the residuals for each recovery. Right panel: The upper panel displays a random SFH draw from each median PSD. The lower panel displays the distribution of SFR$_{\text{10 Myr}}$/SFR$_{\text{100 Myr}}$ as calculated from $N=500$ sampled SFHs from each PSD. A gray dashed line indicates when SFR$_{\text{10 Myr}}$/SFR$_{\text{100 Myr}}$$=$ 1.}
\label{fig:fire_noslope_results}
\end{figure*}

\begin{figure*}
    \centering
    \includegraphics[width=1\linewidth]{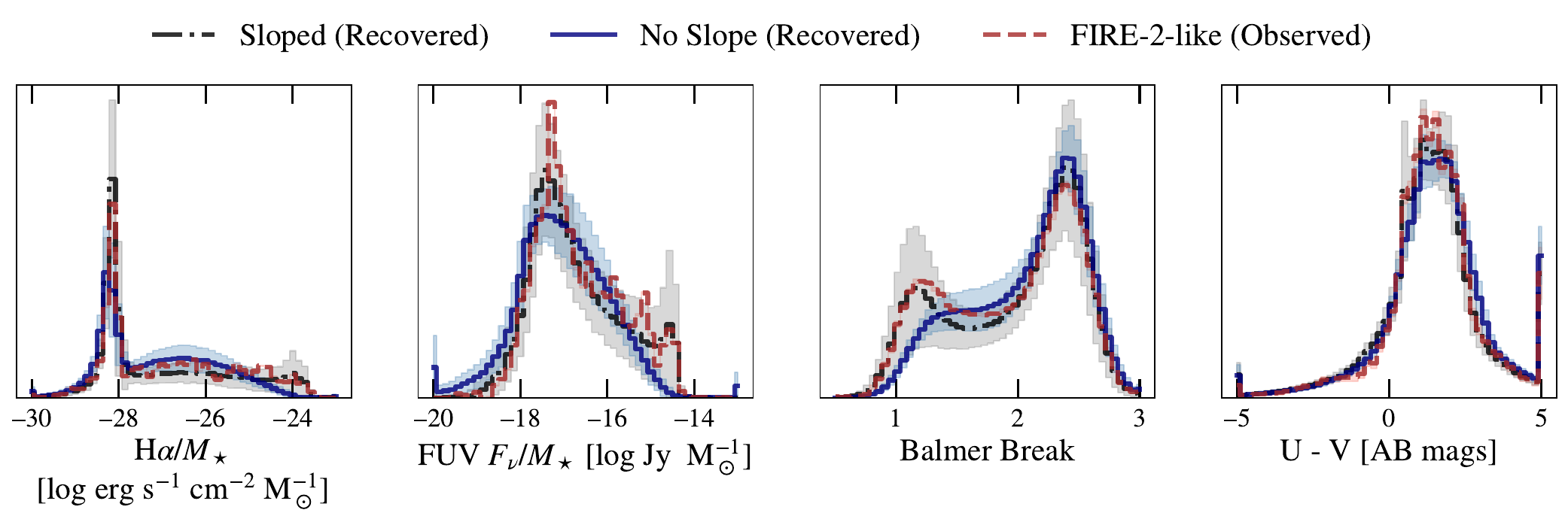}
    \caption{Distribution of observables that were simulated from the FIRE-2-like population PSD (red dashed), and the recovered distributions from the fits to the FIRE-2-like distributions with a model trained without sloped SFHs (blue solid), and with sloped SFHs (black dot-dashed), with their accompanying $\pm1\sigma$ dispersions. The recovered distributions are nearest-neighbor interpolated from each model's training set.}
    \label{fig:fire_noslope_results_dists}
\end{figure*}

The flex-PSD formalism is highly effective for modeling stochastic fluctuations in galaxy SFHs, capturing both bursty and smooth SFR fluctuations across a wide range of timescales (Figure \ref{fig:flexpsd_model}). By describing the variance of $\log\text{SFR}$ as a function of timescale (Equation \ref{eq:PSD}), it provides a quantitative measure of short-term deviations from the long-term average. However, computing the PSD discards all phase information from the original SFH. That is, it characterizes the amplitude of fluctuations without encoding whether a galaxy’s SFR is trending upward or downward over cosmic time. Observationally, galaxy SFHs on a population level are rarely flat, even on average: they tend to rise or fall depending on galaxy mass and redshift, with low-mass galaxies generally exhibiting rising SFHs and higher-mass galaxies showing declining SFHs \citep[e.g.,][]{Cowie1996,Kauffmann2003,Pacifici2013,Behroozi2013,Wang2023:pbeta}. When fitting individual galaxy SEDs with a PSD-based prior, as done in \citet{Wan2025}, this is not as obvious; the inference method would be able to reproduce the observed SED without directly modeling the long-term slope of the SFH. However, when analyzing galaxy \textit{populations}, it is necessary to account not only for stochastic fluctuations but also for the population-averaged SFH slope. Populations with systematically rising or falling SFHs produce observables skewed toward indicators of high or low SFRs (e.g., elevated H$\alpha$ fluxes, weak Balmer breaks, and bluer colors for rising SFHs; see Figure \ref{fig:flexpsd_model_mega}). Combined with the symmetry limitation, this means that the flex-PSD formalism alone cannot capture all aspects of galaxy SFHs, motivating explicit modeling of the population-averaged SFH slope.

In the formalism of Section \ref{sec:psd_model}, the parameters $\mu_\alpha$ and $\sigma_\alpha$ describe the mean shape and scatter of the population’s SFHs, while the PSD parameters characterize stochastic fluctuations about this mean. To illustrate the importance of including the slope parameters, we trained a model using only the flex-PSD parameters, fixing the mean slope and its dispersion to zero ($\mu_\alpha = \sigma_\alpha = 0$). Figure \ref{fig:fire_noslope_results} shows that the inferred PSD (blue line and shaded region) fails to match the FIRE-2-like PSD on timescales longer than 100 Myr within the 68\% posterior confidence interval, indicating extrapolation beyond the training data. Figure \ref{fig:fire_noslope_results_dists} further demonstrates that the mock distributions cannot be reproduced by the model trained without slope parameters. If used as a prior for individual SFH fits, a mischaracterized PSD biases the inferred SFHs to be smooth, as seen in the top right panel of Figure \ref{fig:fire_noslope_results}. Random SFHs sampled from each PSD show that the median FIRE-2-like PSD and the median inferred PSD with slope produce similar short-timescale bursts ($\lesssim 50$ Myr). In contrast, the PSD inferred from the slope-free model produces overly smooth SFHs, with bursts occurring on timescales $\gtrsim 200$ Myr. Consequently, burstiness metrics such as SFR$_\text{10 Myr}$/SFR$_\text{100 Myr}$ are systematically closer to unity (bottom right panel of Figure \ref{fig:fire_noslope_results}). To accurately recover the fluctuations, one must explicitly model the average slope in SFR(t), as these produce similar signatures as burstiness.

\subsection{Populations with Extreme Bursts Obscure Features on Long Timescales}\label{sec:discussion_outshining}

\begin{figure*}
    \centering
    \includegraphics[width=1\linewidth]{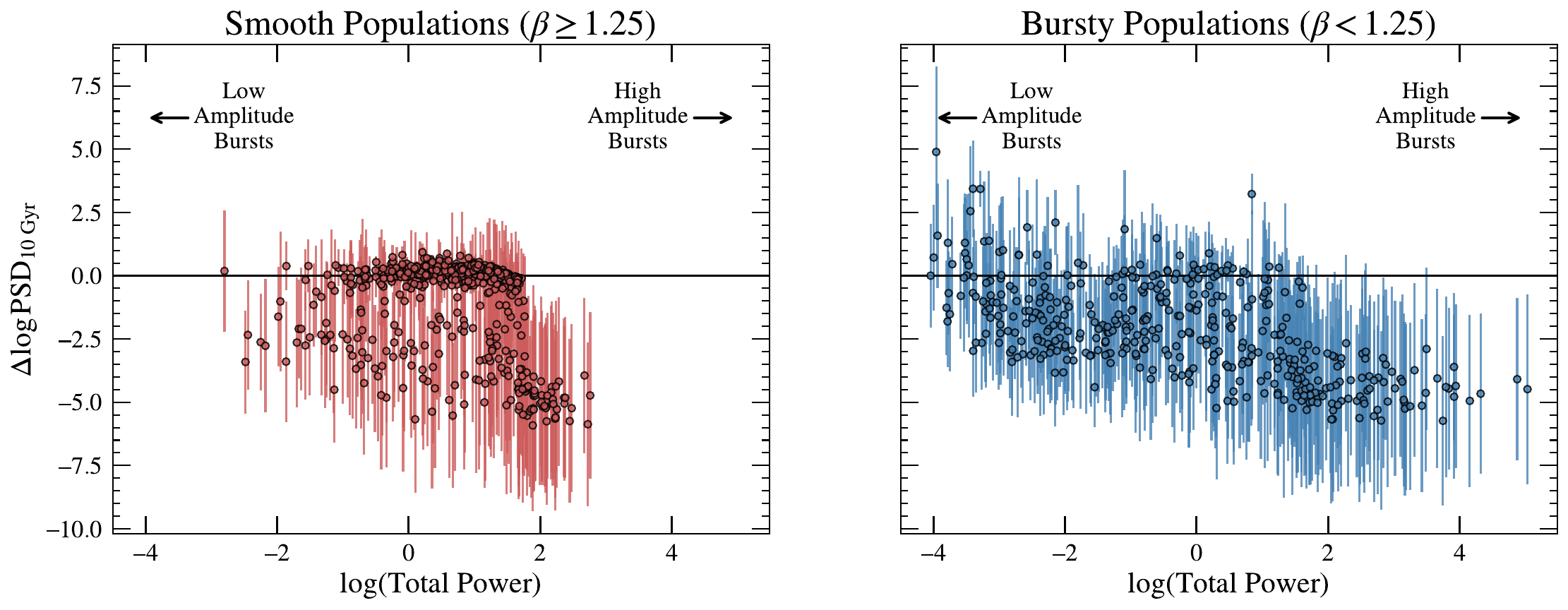}
    \caption{The difference between the median predicted flex-PSD parameter $\pm 1\sigma$ on 10 Gyr timescales versus the total power of the PSD for the ``realistic" test population. Higher total power indicates high amplitude SFR fluctuations, and lower total power indicates low amplitude SFR fluctuations across all timescales. The colors of the scatter points correspond to a best-fit power-law slope, $\beta$, to each PSD, where larger values are considered to be smooth (red, left panel), and smaller values are considered to be bursty (blue, right panel). A solid black line illustrates the dividing line between over- and under-predicting the true value of the flex-PSD parameter.}
    \label{fig:flexpsd_model_realistic_results_correlations}
\end{figure*}

As shown in Sections \ref{sec:results_psd} and \ref{sec:results_outshining}, the power of SFR fluctuations on $\gtrsim 100$ Myr timescales is well constrained in smooth populations similar to Illustris, but is more uncertain in bursty populations like FIRE-2-like. This is evident in the dispersion of the inferred posterior PSD (Figure \ref{fig:fire_illustris_results}), where the 68\% confidence interval is roughly 1 dex wider for the FIRE-2-like population compared to the Illustris-like population, and PSDs inferred from most mock populations show greater uncertainty at long-timescales (Figure \ref{fig:flexpsd_model_realistic_results_psd}). 

Here, we explore the possibility that extremely bursty populations, with significant power on $\lesssim 100$ Myr timescales, can obscure long-timescale fluctuations due to outshining by short-term bursts of star formation. To do so, we inspect how the PSD’s intrinsic shape correlates with the accuracy of its recovery in the ``realistic" galaxy populations discussed in Sections \ref{sec:methods_realistic} and \ref{sec:results_outshining}. This correlation is shown in Figure \ref{fig:flexpsd_model_realistic_results_correlations}. In this figure, the vertical axis shows the difference between the median posterior and the true value of the flex-PSD parameter $\log\mathrm{PSD}_{\text{1 Gyr}}$, while the horizontal axis shows the total power in the PSD ($\int \mathrm{PSD}(f)\,df$). The total power effectively quantifies the overall level of stochasticity in the SFH: very low values correspond to smooth SFHs, whereas higher values reflect increasingly bursty behavior, in the sense that bursts on any timescale become higher in amplitude. The points are colored according to the best-fit single power-law slope of the PSD ($\beta$), where smooth/steep PSDs are shown in red and bursty/shallow PSDs are shown in blue. 

Figure \ref{fig:flexpsd_model_realistic_results_correlations} illustrates that the recovery of long-timescale power worsens for the most bursty populations. These populations, those with large total power and shallow PSD slopes, tend to show increasingly negative residuals, indicating that the median of the posterior systematically under-predicts the power at $\sim$10 Gyr. In such cases, frequent high-amplitude bursts of star formation obscure the weaker long-timescale fluctuations in the observable distributions, making them difficult for the model to recover. This behavior appears regardless of whether the typical SFH of the population is rising or falling.  In contrast, populations with relatively little short-timescale power allow both short- and long-term SFH behavior to be recovered with higher accuracy. Taken together, these results suggest that when short-timescale power is elevated, as found in FIRE-like populations, the power of long-term fluctuations becomes intrinsically difficult to infer, whereas smoother populations permit reliable recovery of the power on both short- and long-timescales. It remains to be seen which regime is most representative of real galaxy populations. Nonetheless, this correlation between burst strengths, timescale, and recovery accuracy may correspond to an underlying limitation in inferring the long-timescale SFH from the integrated light of galaxies, even when using population-level analyses and sophisticated inference techniques.

The constraints on SFH fluctuations are directly linked to the choice of observables, which probe different characteristic timescales (Figure \ref{fig:flexpsd_model_mega}). In this work, we focus on rest-UV to rest-optical indicators that are sensitive primarily to star formation on timescales from a few Myr up to $\sim 1$ Gyr. Because we do not include longer-wavelength observables that trace evolved stellar populations (e.g., rest-frame near-IR colors that probe the red/asymptotic giant branch), our ability to constrain fluctuations on timescales $\gtrsim 1$ Gyr is limited \citep[e.g.,][]{Conroy2013}. Including such longer-timescale indicators may help alleviate the bias introduced by highly bursty populations, as they carry information about older stellar populations that are less affected by short-term bursts. However, we note that current efforts to model the contributions from evolved stellar populations to an integrated galaxy SED remain uncertain, and thus caution is warranted when interpreting constraints derived from such observables \citep[e.g.,][]{Kriek2010,Walcher2011}.

\subsection{Systematics of SPS and Dust Attenuation}\label{sec:discussion_systematics}

A core assumption of this method is that the chosen SPS model (which encompasses the stellar spectral templates, isochrone library, IMF, stellar metallicity distribution, dust treatment, alpha element enhancement, modeling of evolved stars, and related assumptions) is representative of the galaxy population on which the trained model is evaluated. If these assumptions are mismatched, the distribution of timescale-sensitive features can change, potentially biasing the inferred population parameters. This issue has long challenged efforts to constrain burstiness in galaxy populations \citep[e.g.,][]{Emami2019}. For example, a model trained on observables derived from the MILES spectral library but evaluated on populations with a high binary fraction that more closely resemble BPASS \citep{Eldridge2009} templates may yield systematically different features, particularly for short-timescale indicators such as \Halpha. In principle, it may be possible to train a model that marginalizes over the uncertainty introduced by different stellar template choices. We note, however, that the effects of variance in stellar population templates, especially ionizing photon production efficiencies, have not yet been explored in this work. The efficacy of the method, therefore, depends on selecting stellar templates appropriate for the target population.

The chosen metallicity distribution for the population is also a significant systematic. In this work, we adopted a simple uniform stellar metallicity distribution. It would be entirely feasible for the metallicity distribution, or parameters that describe said distribution (e.g., a mean and dispersion of the stellar metallicity), to also be population-level parameters that are fitted for alongside the SFH burstiness parameters. On the other hand, the distribution may be held fixed, or perhaps directly correspond to the mass-metallicity relation \citep[e.g.,][]{Tremonti2004}. Accounting for metallicity in this way has the potential to improve constraints on SFH stochasticity, particularly if the metallicity distribution is tied to variations on specific timescales.

Dust attenuation is another important systematic, as dust strongly affects the observed galaxy SED \citep[e.g.,][]{Calzetti2001}. Attenuation is typically a strong function of wavelength and may correlate with intrinsic galaxy properties such as SFR and stellar mass \citep[e.g.,][]{Meurer1999,Reddy2015}. The observables considered in this work exhibit differing sensitivity to dust: rest-frame UV-optical colors are the most strongly affected, followed by continuum break strengths and emission/absorption lines. When multiple Balmer lines are observed in emission, this effect can be partially mitigated using the Balmer decrement, although it still requires a choice of parametrization for the attenuation curve. Using a set of observables without properly accounting for dust is likely to produce unreliable results, as a model trained on dust-free galaxy populations would be extrapolating beyond its training set. In principle, the method could be extended to include one or more population-level parameters describing the dust attenuation curve, allowing the model to marginalize over dust properties simultaneously with the population-level SFH parameters. However, this would likely require an increase in training set size. As an intermediate step, one could inflate the observational uncertainties by factors chosen to approximate dust-correction uncertainties and test whether the resulting population model retains useful constraining power. We leave a quantitative exploration of these approaches to future work.

\subsection{Practical Considerations in Applying the Method to Spectroscopic Samples}\label{sec:discussion_obs}

We now conclude this section by discussing how our method can be applied to observations. The methodology introduced in this work is designed for application to statistical, representative spectroscopic samples of galaxy populations. In practice, utilizing this framework involves a sequence of choices: defining a population, selecting observables, specifying a physical model, and finally evaluating the SFH PSD.

First, the galaxy population must be defined. In practice, users should group galaxies according to the factors they believe most strongly drive burstiness, depending on the science question being addressed by the population. Such criteria for defining a population could include a relatively narrow redshift distribution and/or comparable stellar masses. Star formation timescales are expected to evolve with cosmic time through changes in global gas fractions \citep{Madau2014}. As a result, higher gas fractions at early epochs naturally produce higher SFR surface densities via the Kennicutt-Schmidt relation \citep{Schmidt1959, Kennicutt1989}. A galaxy's dynamical mass correlates with the dominant feedback processes that regulate short-timescale SFR fluctuations, a trend explicitly seen in the PSDs of simulated SFHs: lower-mass galaxies ($M_\star < 10^{10} \ \mathrm{M_\odot}$) exhibit systematically higher power of at $\lesssim 300$ Myr timescales compared to higher-mass systems in most of the hydrodynamical simulations investigated by \citet{Iyer2020}. Together, these criteria suggest the method performs best for populations in which the underlying physics governing star formation are broadly similar (e.g., AGN presence, dynamical structure, environment, age). In practice, this information may come from existing photometry, prior stellar mass and redshift estimates, or even post-hoc grouping from a spectroscopic sample. Ultimately, regardless of the specific criteria used to define galaxy populations, the method constrains the population-averaged characteristics of their SFHs. Even if individual galaxies differ in detail or the grouping is not perfect, the inferred PSDs provide insight into the timescales of burstiness across the population.

Second, the timescale-sensitive observables are chosen. Because the inference performance relies on the information content of the observable distributions, users may tailor the choice of observables to target specific SFH timescales. Beyond the indicators explored in this work, additional constraints may be gained from other Balmer emission lines, rest-UV to rest-NIR colors, and, where signal-to-noise permits, stellar absorption features such as Ca H \& K. For best results, observables should be selected to collectively probe a wide range of timescales, from short-lived massive stars to more evolved stellar populations.

Third, a physical SPS model is specified. Once a population and set of observables are defined, applying this method requires selecting a physical model appropriate for the galaxy population of interest and specific science goal(s), as described in Section \ref{sec:methods_sps}. This includes defining the distributions of nuisance parameters to marginalize over (e.g., stellar metallicity, dust), selecting a population-level SFH model, measurement noise models, and specifying the configuration used to train the SBI model. This analysis can be repeated for different subgroups within a population to assess the robustness of the results. 

Finally, one is able to evaluate a model trained on the observables generated by the physical model chosen to infer the PSD. Constraining the SFH PSD of a galaxy population also enables several avenues for its interpretation. The inferred PSD can serve as a prior for the SFH in subsequent SED fits of individual galaxies belonging to said population. Population PSDs additionally offer a direct interface with cosmological simulations: comparing observed PSDs to simulated ones yields quantitative tests of feedback prescriptions and the characteristic timescales of SFHs. Additionally, PSD measurements across different redshifts or environments allow evolutionary studies of SFR timescales, potentially revealing how the dominant feedback processes evolve over cosmic time and how they may correlate with the physical properties of galaxy populations. 

\section{Conclusions}\label{sec:conclusions}

In this work, we introduce a population-level inference framework that infers the PSD describing galaxy populations’ SFHs from the marginal distributions of timescale-sensitive spectral features. This forward-modeling approach enables quantitative constraints on both short- and long-timescale fluctuations in galaxy populations’ SFHs.
Our main conclusions are as follows.

First, when applied to an illustrative oscillating SFH parametrization as a proof-of-concept, we find that the burst amplitudes and timescales can be constrained to within $<0.1$ dex and $\lesssim3$ Myr, respectively, with well-calibrated posteriors.

Second, we demonstrate that the power of SFR fluctuations can be inferred within the 95\% confidence intervals of the posteriors of flex-PSD model parameters for both smooth and bursty SFH populations across 1 Myr-10 Gyr timescales. We find that the structure of the posteriors is often non-Gaussian, and that the posterior median is likely to underestimate the power on $1-10$ Gyr timescales.

Third, we show that modeling the average SFH slope of a population is essential for correctly interpreting the distributions of timescale-sensitive observables in predominantly star-forming or quiescent populations. Neglecting the mean SFH slope in the training set leads to systematically underestimated burstiness ($\sim$1 dex less power on 1-10 Myr timescales and 0.5 dex more power on 1 Gyr timescales) and overly smooth inferred SFHs. These results imply that burstiness estimates derived from simple diagnostics that do not control for secular SFH trends are likely biased toward underestimating the amount of burstiness.

Fourth, our method successfully recovers PSDs for populations with smooth to moderately bursty SFHs. For populations experiencing intense (high amplitude) and frequent (on $\lesssim 10$ Myr timescales) bursts, observable features of long-timescale fluctuations are obscured, increasing the uncertainty of the inferred power on these timescales. This potentially reveals a fundamental observational limit of constraining long-timescale variations from the integrated light of galaxies due to the outshining of young stellar populations.

Finally, applying this framework to simulations shows that it can recover the short-timescale ($\lesssim 100$ Myr) SFR fluctuation power to within $\sim$1 dex, allowing FIRE-2-like and Illustris-like feedback prescriptions to be distinguished with $>99$\% confidence, and thereby providing direct observational constraints on theoretical models of feedback mechanisms that drive star formation.

Open avenues with this method include the addition of dust to the physical model, constraining the robustness of this method when applied to incomplete spectroscopic samples, samples with known selection functions, and quantifying the systematics between SPS models. Equally important is determining how constraints on the population-level PSD influence the inference of key physical properties of galaxy populations, such as stellar masses, the dispersion in the SFMS, and SFRs. We leave the exploration of these topics to future work.

With JWST entering its fourth year of operation, large, statistically representative samples of galaxy populations in the distant Universe are now being assembled. The method introduced in this work is designed to constrain the strength and timescales of the feedback driving star formation across such populations, providing a direct link between observations and the physical mechanisms regulating galaxy evolution.

\section*{Acknowledgments}
E.B. acknowledges support through program JWST-GO-04111.014-A, which was provided through a grant from the Space Telescope Science Institute under NASA contract NAS 5-03127.
B.W. acknowledges support provided by NASA through Hubble Fellowship grant HST-HF2-51592.001 awarded by the Space Telescope Science Institute, which is operated by the Association of Universities for Research in Astronomy, In., for NASA, under the contract NAS 5-26555. 
S.W. received support from NASA grant 80NSSC24K0838.
T.B.M. was supported by a CIERA fellowship.
Computations for this research were performed on the Pennsylvania State University’s Institute for Computational and Data Sciences’ Roar supercomputer. This publication made use of the NASA Astrophysical Data System for bibliographic information.

This work is partly based on observations from the NASA/ESA/CSA James Webb Space Telescope.
Data was accessed via the JADES public data release 3 emission line catalog, which can be viewed at \hyperlink{https://archive.stsci.edu/doi/resolve/resolve.html?doi=10.17909/8tdj-8n28}{10.17909/8tdj-8n28} and the UNCOVER program public data release 3 SUPER photometric catalog, which can be accessed at \hyperlink{https://doi.org/10.5281/zenodo.10987924}{10.5281/zenodo.10987924}.

\textit{Software:}
\texttt{FSPS} \citep{Conroy2009,Conroy2010},
\texttt{matplotlib} \citep{matplotlib2007},
\texttt{numpy} \citep{numpy2020},
\texttt{sbi} \citep{TejeroCantero2020},
\texttt{scipy} \citep{2020SciPy},
\texttt{sedpy} \citep{Johnson2021_sedpy},
\texttt{Prospector} \citep{Leja2017,Johnson2021_prospect}, and
\texttt{python-fsps} \citep{Johnson2024_pythonfsps}.

\bibliography{timescales}{}
\bibliographystyle{aasjournal}

\appendix

\section{Inferring parameters from Joint vs. Marginal Observable Distributions}\label{app:joint}

\begin{figure*}[htbp]
\centering
\gridline{
    \fig{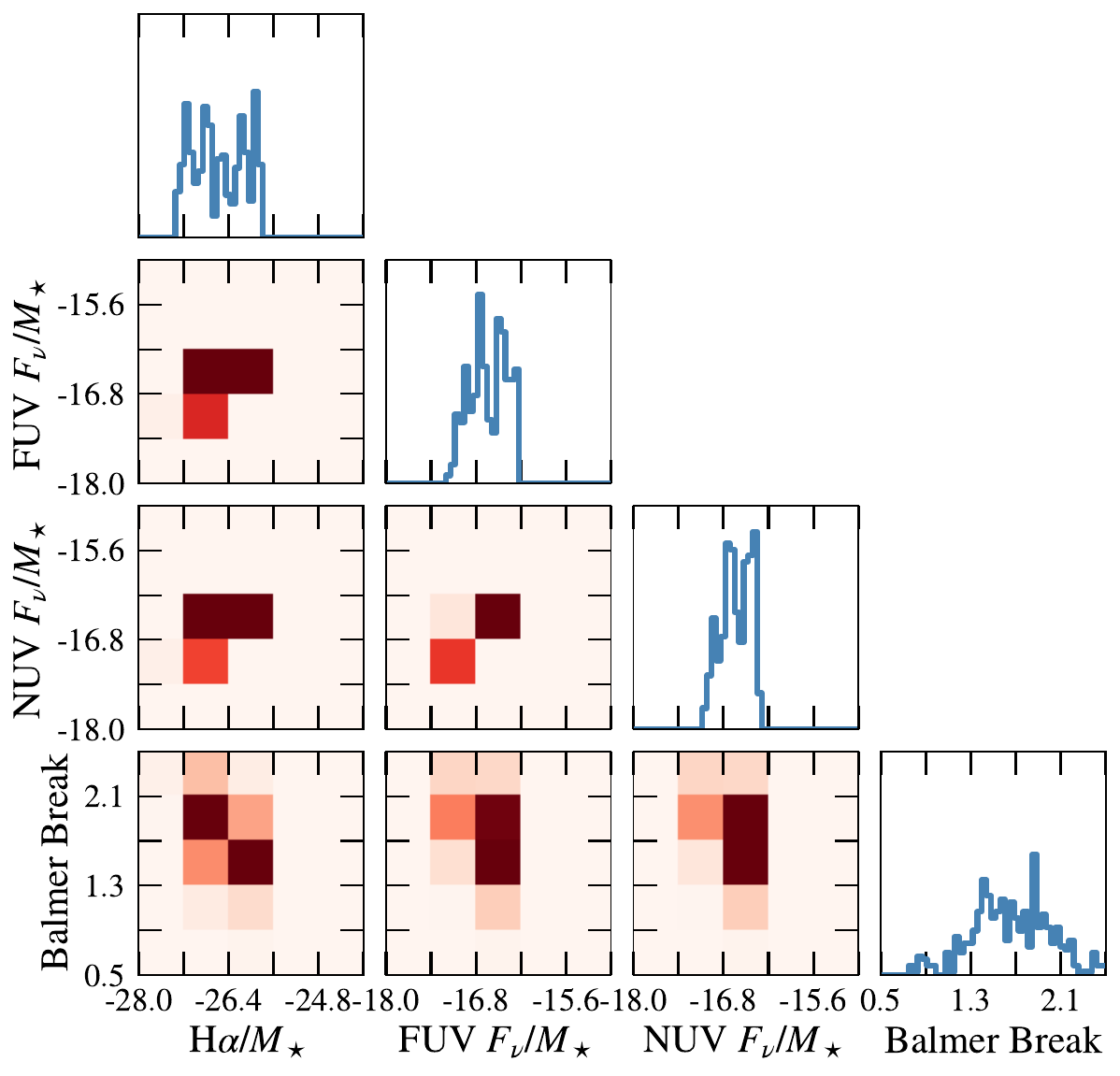}{0.49\textwidth}{(a)}
    \fig{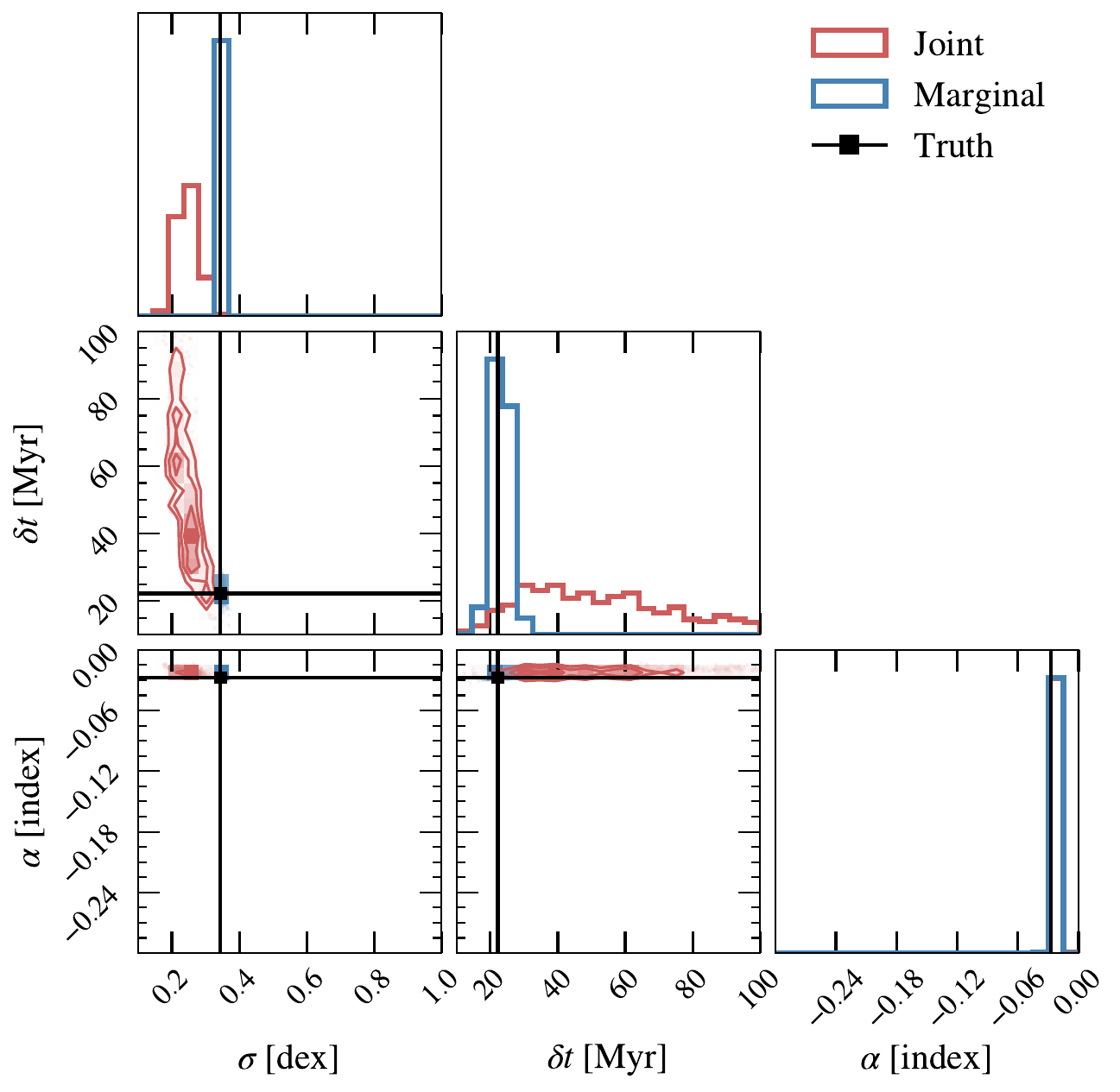}{0.49\textwidth}{(b)}
}
\caption{Recovery of parameters for single-frequency SFH model using marginal versus the joint distribution of observable features. 
Left panel: Input features for the SBI models, high-resolution marginal distributions (blue, 50 bins per observable), and coarse joint distributions (red, 5 bins per observable). 
Right panel: Resulting posteriors of the single-frequency SFH model parameters inferred from the distributions in the left panel. The true parameter values are displayed in black.}
\label{fig:marginal_joint}
\end{figure*}

This work has only inferred parameters from the marginal (i.e., one-dimensional) distribution of features in a population's spectra. In this section, we explore utilizing the joint (i.e., multi-dimensional) distribution of the observable features of galaxy populations as inputs to the SBI model. 

We take the training set of the single-frequency SFH model, as discussed in Section \ref{sec:simple_model}, and instead calculate multivariate histograms of the observables in the simulated populations via \texttt{numpy.histogramdd} with 5 bins per observable. This results in 625 total input features to the model for four spectral features, as compared to the 200 input features for four spectral features with 50 bins each for the univariate histograms. These histograms are flattened, artificial noise is added, and an SBI model is trained in the same manner discussed in Sections \ref{sec:methods_dists} and \ref{sec:methods_sbitraining}. 

The left panel of Figure \ref{fig:marginal_joint} displays the two-dimensional representation of the multidimensional inputs to the model for an example mock population with the single-frequency SFH parameters in red shades. The marginal distributions are displayed in blue for visualization purposes. When the trained model is evaluated, given the joint distributions, the red contoured posterior samples displayed in the right panel of Figure \ref{fig:marginal_joint} are inferred. The population parameters are recovered within $1\sigma$ given the coarse, joint distribution of observables. However, the posteriors are wide, indicating a greater level of posterior uncertainty. Contrastingly, the model trained on the fine, marginal distribution of the observables yields a posterior with high confidence and precisely predicts the population parameters from the distributions. Notably, the joint instance has a $\sigma_\text{NMAD}$ that is increased by a factor of $\times 2$, but the model uncertainties remain well-calibrated for both models.

This stark difference in the inferred posteriors for the same population, despite identical underlying observables that were merely reformatted prior to model evaluation, suggests that the performance disparity arises from differences in the effective information content (or resolution) of the observable distributions. Because the same simulated populations, noise realizations, and SBI training procedure are used in both cases, the binning scheme is the only factor that differs between the two models. Although the neural networks within the SBI model receive a larger number of input features, the observables themselves are sampled at lower intrinsic resolution, limiting their ability to distinguish subtle variations in the population SFH model. As a result, the model cannot reliably capture small-scale structure or correlations present in the higher-resolution formulation of the same data.

In principle, the joint distributions are expected to provide stronger constraints on the inferred timescales, but the effective resolution of this joint information is a crucial consideration. When the population size is fixed, coarse joint binning can therefore reduce, rather than enhance, the constraining power of the observables. We conclude that histogram density values are not efficient summary statistics of high-dimensional information for inference purposes. As an alternative, one may adopt a non-parametric representation of the population (e.g., summary statistics from the spectra themselves, or the distribution of timescale-sensitive indicators) in place of a histogram. Possible implementations include a Gaussian mixture model decomposition, a convolutional/embedding neural network, or image moments. We leave the validation of this adjustment of the method to future exploration.

\section{Population Sample Size Directly Impacts Model Parameter Recovery}\label{app:population_size}

\begin{figure*}
    \centering
    \includegraphics[width=1\linewidth]{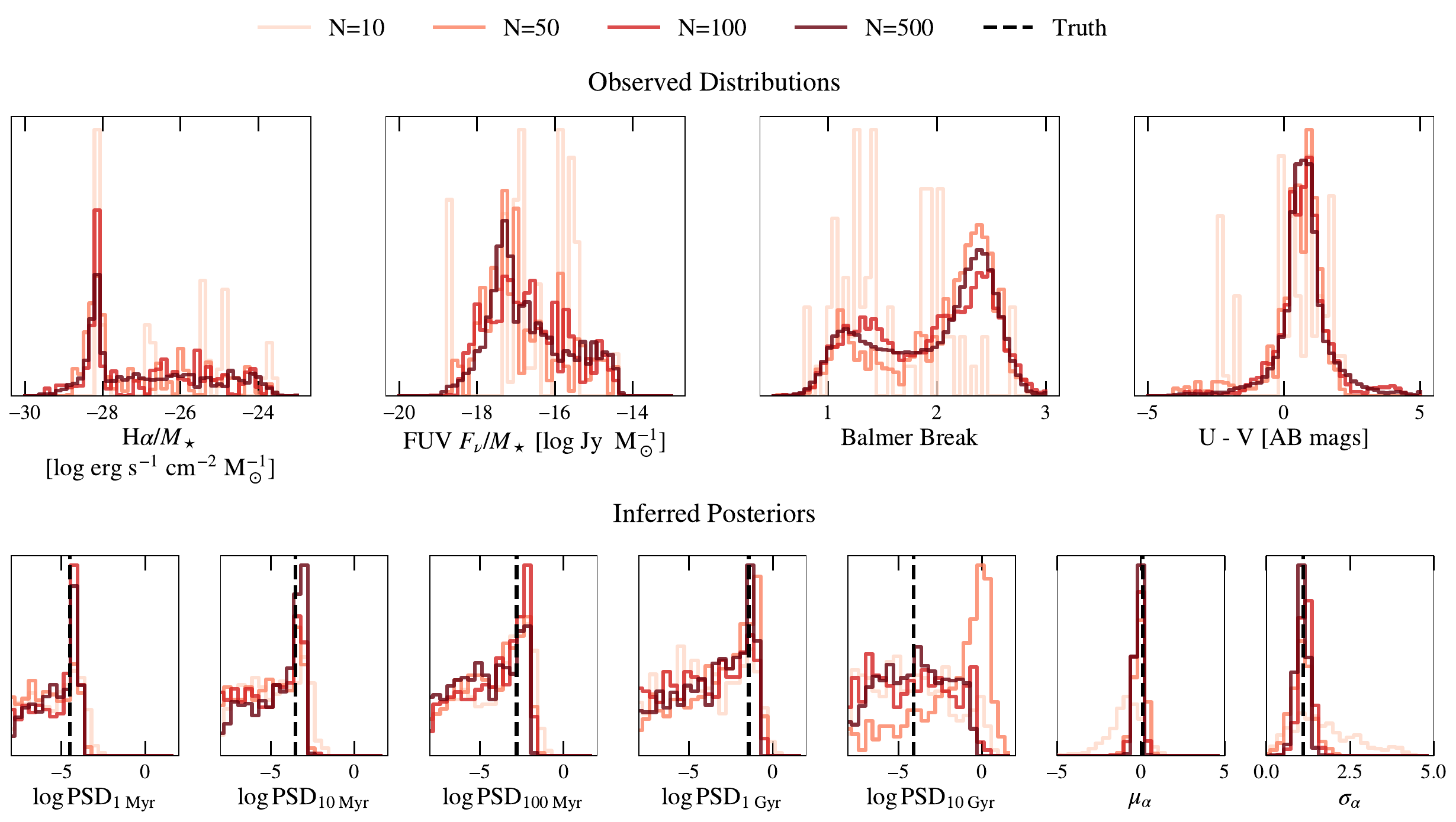}
    \caption{Small population sample sizes lead to poor recovery of population-level parameters. Upper row: distributions of mock observables in FIRE-2-like galaxy populations with different sample sizes. We vary the population size, $N$, to be $N=$ 500 (dark red), 100 (red), 50 (light red), and 10 (light pink). Lower row: Inferred marginal posteriors for each of the population-level parameters, five flex-PSD parameters, and two SFH slope parameters. These posteriors are the result of an SBI model trained on a specific population sample size, which was evaluated on the distributions from the top row's panels.}
    \label{fig:population_size}
\end{figure*}

To investigate how the galaxy population sample size impacts parameter recovery in this method, we train three additional SBI models on the flex-PSD and sloped SFH population parameters, varying only the population size from the fiducial $N=500$ to $N=100$, $50$, and $10$. The results of evaluating each model on mock FIRE-2-like populations are shown in Figure~\ref{fig:population_size}. The top row displays the observed mock distributions passed to the model, while the lower panels show the inferred marginal posteriors for each of the seven population-level parameters.

We interpret the degradation in parameter recovery, for this specific binning scheme, as a consequence of increased counting uncertainty dominating the noise budget of the observed distributions. Indeed, both the observed distributions and the inferred marginal posteriors in Figure \ref{fig:population_size} are remarkably similar for population sizes of $N=100$ and $N=500$, indicating that measurement uncertainty dominates in these regimes. However, when decreasing the sample size from $N=100$ to $N=50$, increased counting noise becomes apparent in the observed distributions, and the marginal posteriors broaden accordingly, reflecting decreased inference confidence. For $N=10$, there is effectively insufficient information to constrain the PSD parameters, although the SFH slope parameters remain weakly constrained with large uncertainties. When projected into the recovery of the PSD, we find that fewer than $N=100$ galaxies in a population provide distributions dominated by counting noise, and thus the model returns a flat PSD, which is the prior distribution for the training set.

These results demonstrate that the model remains well-calibrated even when trained and evaluated on low-sample-size populations, but that robust constraints on the characteristic timescales require at least $N \gtrsim 100$ galaxies per population for this summary statistic scheme. This is approximately the number of spectra that can be obtained with a medium-sized JWST observing program. A complementary trade-off must also be considered: decreasing bin resolution to decrease the impact of counting uncertainty similarly broadens the inferred posteriors, as discussed in Appendix \ref{app:joint}. 

\section{Motivation for a Flexible Population-Level PSD Model}\label{sec:app_exreg}

A physically-motivated formalism to model stochastic SFHs is the Extended Regulator Model, as introduced in \citet{Tachella2020}. 
The Extended Regulator Model quantifies the strength of stochastic star-formation fluctuations in the form of a broken power-law PSD curve over some set of frequencies $f$:

\begin{equation} \label{eq:ExReg}
    \text{PSD}_{\text{ExReg}} = s^2_{\text{reg}} \left(  \frac{1}{1 + (2 \pi \tau_{\text{in}})^2 f^2} \times \frac{1}{1 + (2 \pi \tau_{\text{eq}})^2 f^2}  \right) + \frac{s^2_{\text{dyn}}}{1 + (2 \pi \tau_{\text{dyn}})^2 f^2}
\end{equation} 

\noindent where $\tau_{\text{in}}$, $\tau_{\text{eq}}$, and $\tau_{\text{dyn}}$ are the timescales associated with gas inflow, equilibrium gas cycling, and giant molecular cloud formation/disruption, respectively. $s_{\text{reg}}$ is the power of variability associated with the gas inflow and equilibrium cycling within the population, and $s_{\text{dyn}}$ is the power of the variability in dynamical processes where $s_{\text{reg}} = 2\sigma_{\text{reg}}^2(\tau_{\text{in}} + \tau_{\text{eq}} )$ and $s_{\text{dyn}} = 2\sigma_{\text{dyn}}^2\tau_{\text{dyn}}$ over some timescale $\tau$. This formalism provides five parameters that the methodology of Section \ref{sec:methods_sbi} can easily be adapted to perform population-level inference with. Additionally, the Extended Regulator Model has already been adapted for use in the \texttt{Prospector} pipeline through the work of \citet{Wan2024} in the form of a stochastic prior for non-parametric SFHs. 
Unlike the flex-PSD model, the ACF of the Extended Regulator Model has an analytical solution of the form:

\begin{equation}\label{eq:ACF_ExReg}
    \text{ACF}_{\text{ExReg}} = \sigma_{\text{reg}}^2 \times \frac{\tau_{\text{in}}e^{-|\tau|/\tau_{\text{in}}} - \tau_{\text{eq}}e^{-|\tau|/\tau_{\text{eq}}}}{\tau_{\text{in}}-\tau_{\text{eq}}} \\
    + \sigma_{\text{dyn}}\times e^{-|\tau|/\tau_{\text{dyn}}}
\end{equation}

\begin{table*}
    \centering
    \begin{tabular}{lll} \hline \hline
        \textbf{Parameter} & \textbf{Description} & \textbf{Training Distribution}\\ \hline
         $\sigma_\text{reg} $ & Gas regulation variation (dex) & Uniform [0.1,5.0]\\
          $\tau_\text{eq}$ & Gas equilibrium timescale (Gyr) & Uniform [0.001,1.5]\\
         $\sigma_\text{dyn}$ & Gas dynamical variation (dex) & Uniform [0.01,1.0]\\
         $\tau_\text{dyn}$& Gas dynamical timescale (Gyr) & Uniform [0.001,2.0]\\
         $\tau_\text{in}$& Gas infall timescale (Gyr) & Uniform [0.001,1.5]\\ 
         $\mu_\alpha$ & Mean SFH Slope (power law index) & Uniform [-5,5]\\
         $\sigma_\alpha$ & Dispersion of SFH Slope (power law index) & Uniform [0,5]\\\hline
    \end{tabular}
    \caption{Description of Extended Regulator Model parameters and SFH slope parameters training distributions used to create training set. The distributions of the PSD parameters were chosen to encompass the dynamic range of the prior distributions recommended by \citet{Wan2024}.}
    \label{tab:ExReg_training_distributions}
\end{table*}

\begin{figure}[htbp]
\centering
\gridline{
    \fig{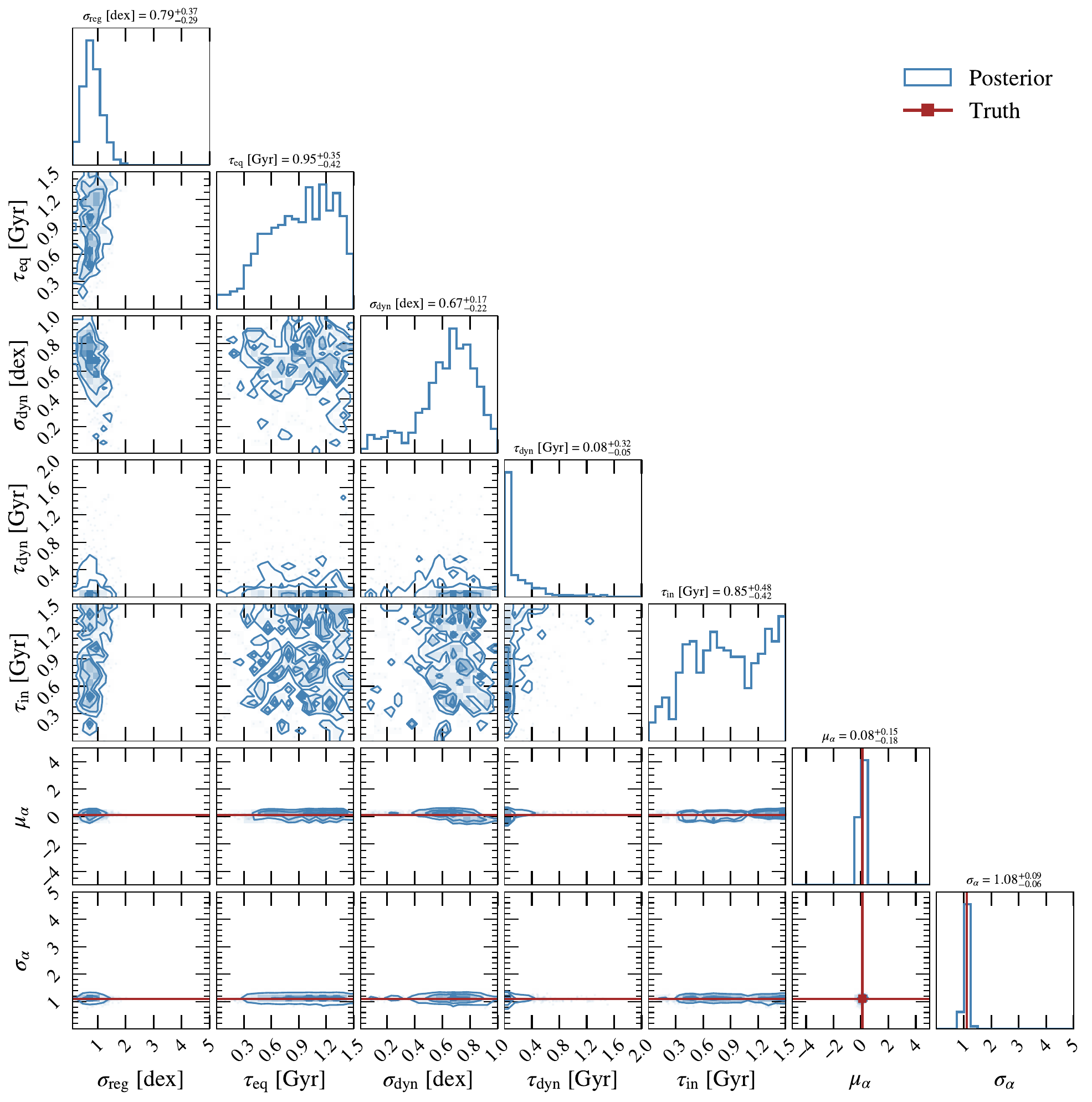}{0.49\textwidth}{(a)}
    \fig{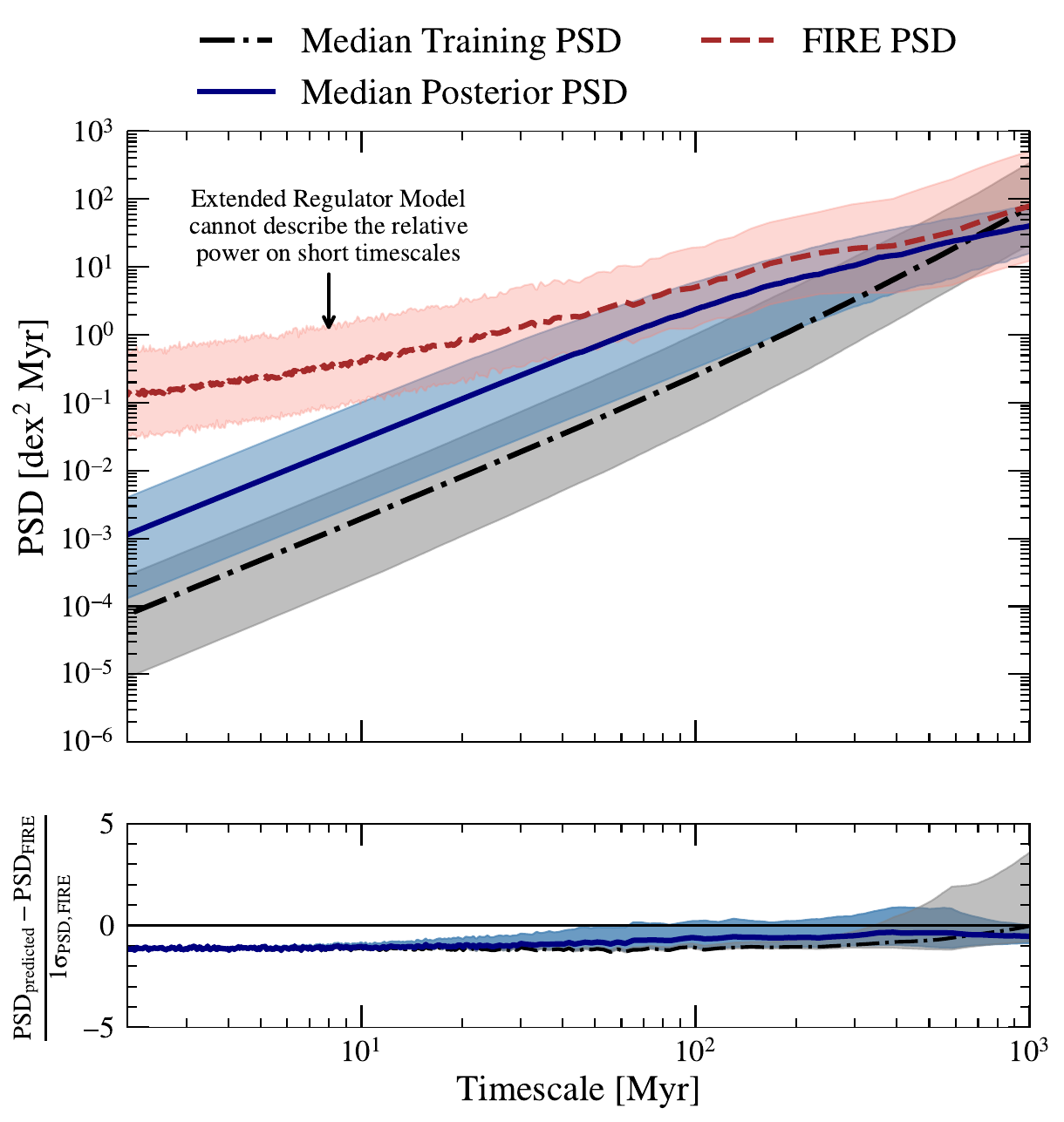}{0.45\textwidth}{(b)}
}
\caption{Recovery of the FIRE-2-like PSD with an SBI model trained on the Extended Regulator model. Left panel: Posterior distribution of the inferred FIRE-2-like Extended Regulator model parameters and population SFH slope parameters. The true value for the slope parameters is shown in red. Right panel: The upper panel displays the posterior median PSD (solid blue) and the training set median (black dot-dashed) compared to the FIRE-2-like population PSD (red dashed). The shaded regions correspond to the 68\% intervals for each set of PSDs. The lower panel displays the residuals of the posterior and training set PSDs versus the FIRE-2-like PSD.}
\label{fig:fire_exreg_results}
\end{figure}

We train an SBI model to recover the five Extended Regulator Model PSD parameters and the two SFH slope parameters with the method of Section \ref{sec:methods_sbi}. The Extended Regulator Model parameter training distributions are provided in Table \ref{tab:ExReg_training_distributions}. We display the recovery of the FIRE-2-like PSD in Figure \ref{fig:fire_exreg_results}, where it can be seen that the Extended Regulator Model trained SBI model cannot correctly infer the shape of the FIRE-2-like PSD on timescales shorter than 10 Myr, as evidenced by the offset between the blue effective posterior PSD and the FIRE-2-like PSD in the right-hand panel.  This is due to the limited training distribution of the model, shown as the gray shaded region.  In other words, the Extended Regulator Model is not able to describe the relative power on short ($<100$ Myr) timescales that is seen in FIRE-2-like.  It is for this reason that we chose to pursue a more flexible PSD model that can easily describe the shape of the FIRE-2-like PSD: the flex-PSD model.

\section{flex-PSD Testing Set Recovery}\label{sec:app_testset}

\begin{figure*}[htbp]
\centering
\gridline{
    \fig{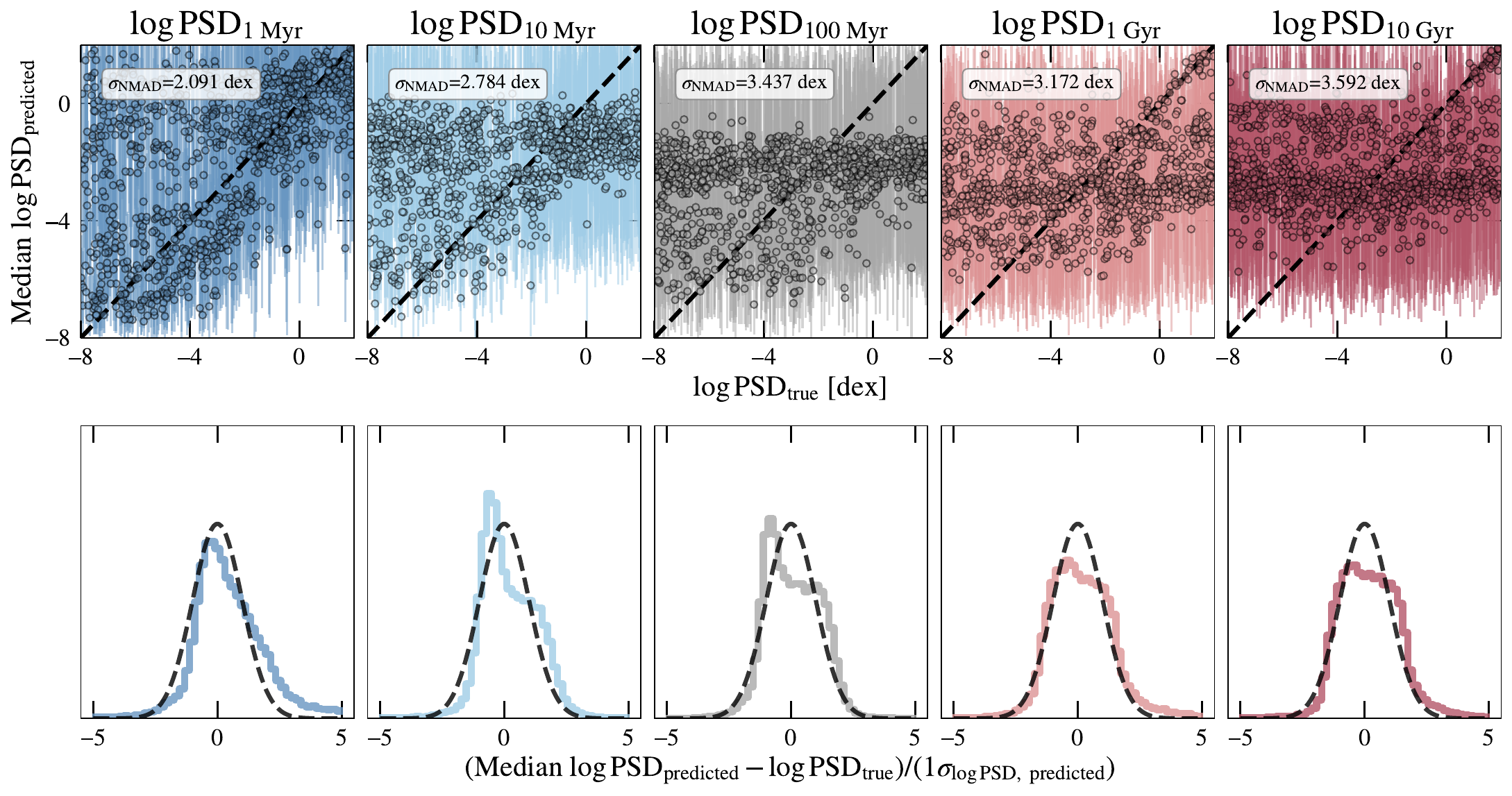}{1.0\textwidth}{(a) SBI model residuals for the flex-PSD parameters.}
}
\gridline{
    \fig{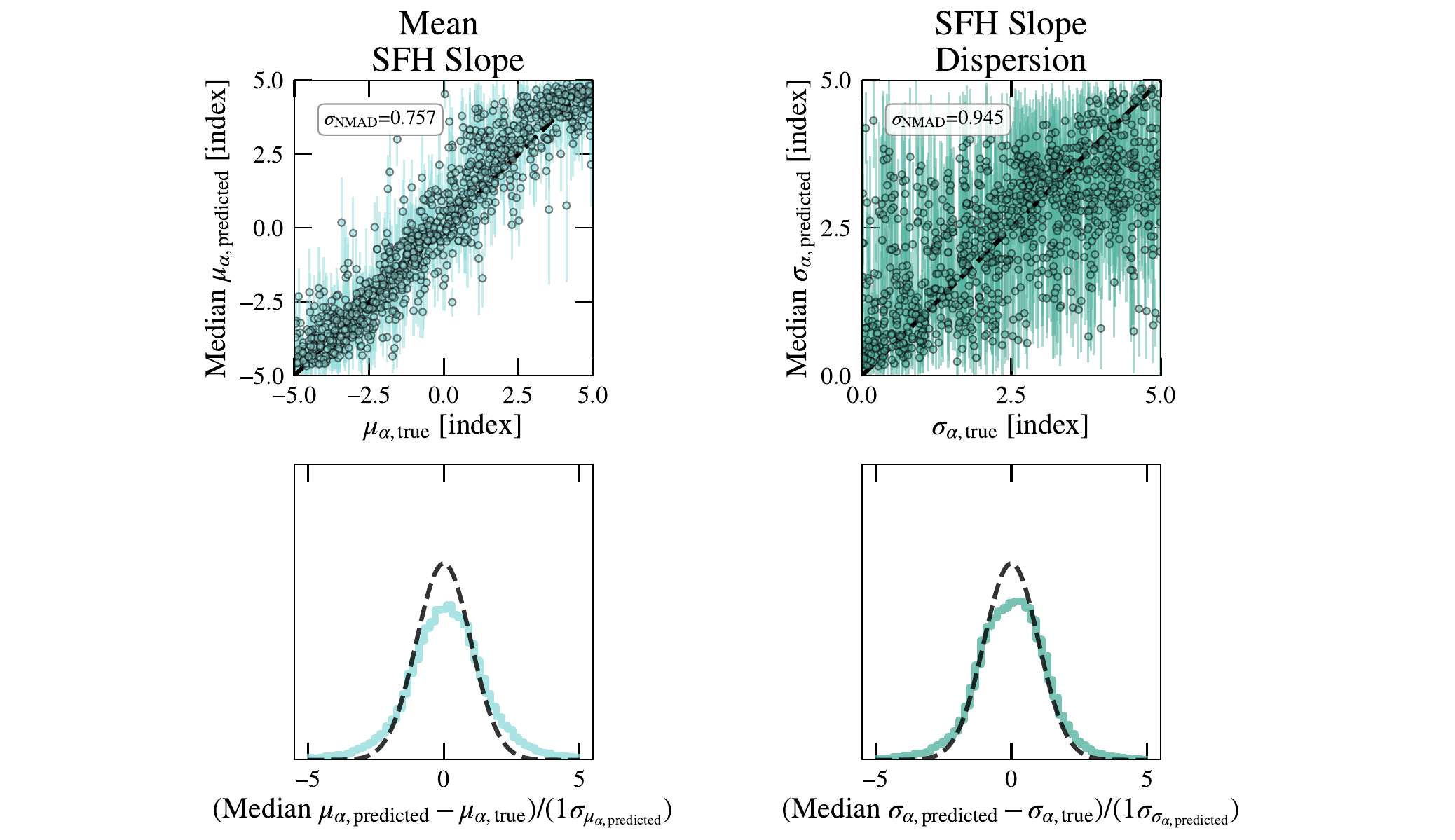}{1.0\textwidth}{(b) SBI model residuals for the population SFH slope parameters.}
}
\caption{Residuals for the flex-PSD population model parameters across the testing set, which is sampled uniformly and independently without informed or correlated priors. In both panels, the upper row displays the median posterior $\pm 1\sigma$ versus the true value of the hyperparameter; a one-to-one line is displayed as a black dashed line for reference. The lower row displays the standardized residuals for each hyperparameter. A unit-Gaussian is displayed as a black dashed line for reference.}
\label{fig:flexpsd_model_results}
\end{figure*}

\begin{figure}
    \centering
    \includegraphics[width=0.5\linewidth]{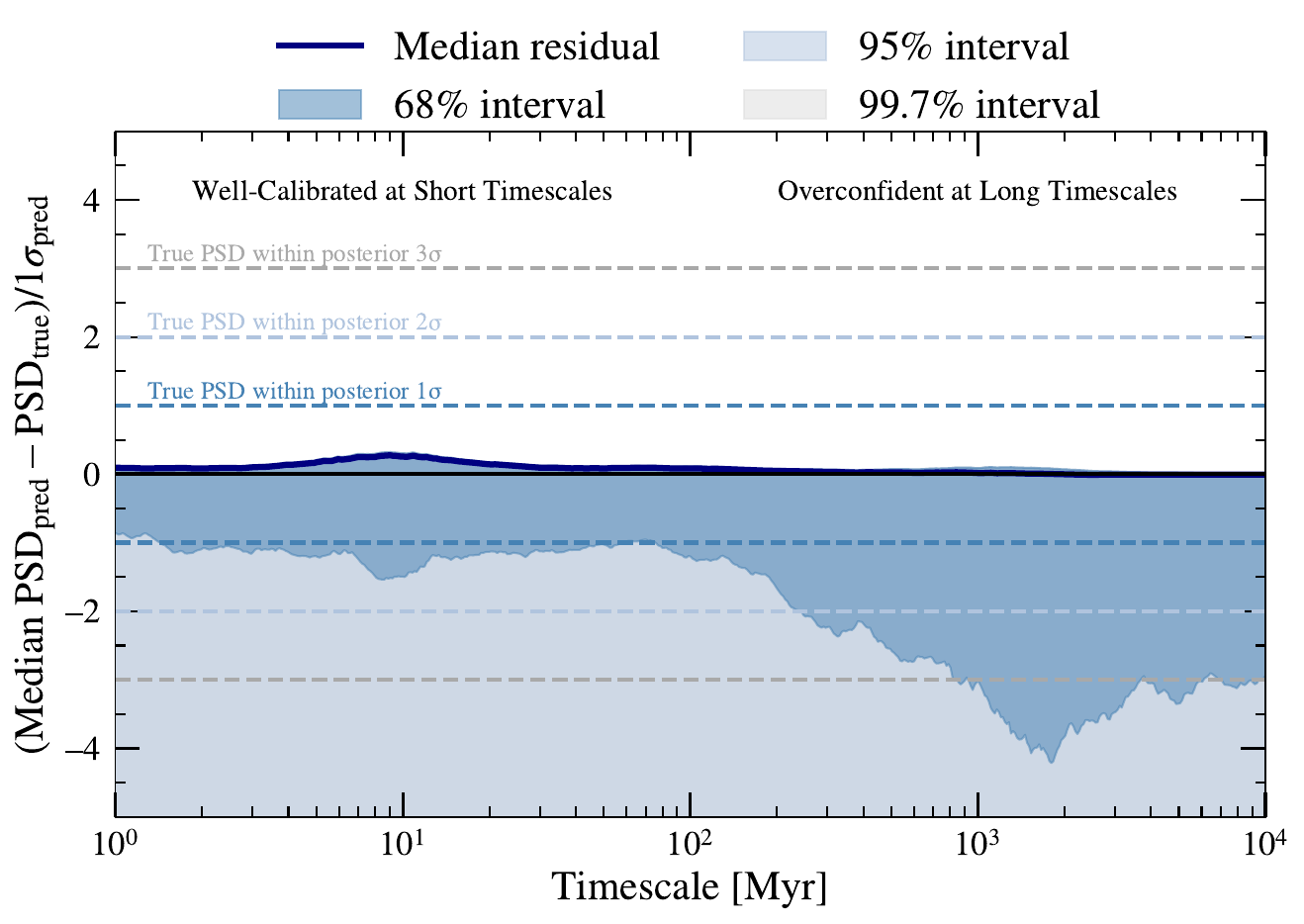}
    \caption{Residuals for the inferred PSD from the testing set flex-PSD population. We display the median residual (dark blue line) as a function of timescale for mock populations' PSDs. The shaded regions correspond to 68\%, 95\%, and 99.7\% intervals for the standardized residuals (blue, light blue, and gray shades). The blue, light blue, and gray dashed lines bracket the area that indicates that the true PSD is within $1\sigma$, $2\sigma$, and $3\sigma$  dispersions of the posterior PSD, respectively.}
    \label{fig:flexpsd_model_psd_results}
\end{figure}

In Section \ref{sec:discussion_outshining}, we discuss the SBI model performance on some test set of ``realistic" SFHs. It is important to note that the SBI model is trained on flex-PSD parameters that are independently and uniformly sampled, so the results discussed are not necessarily representative of the entire parameter space that the model is trained to infer. This section discusses the performance of the model on mock populations that cover the full dynamic range of its training distribution.

Figure \ref{fig:flexpsd_model_results} shows the residuals for a test set of 10$^5$ mock populations sampled independently and uniformly from the training distributions of all parameters. The scatter in the residuals is significantly larger for these “non-realistic” populations, with $\sigma_\text{NMAD}$ increased by approximately a factor of $\times 2$ compared to the results in Figure \ref{fig:flexpsd_model_realistic_results}. However, the model’s predicted uncertainties are notably better calibrated in this case, where the distribution of standardized residuals more closely follows a normal distribution. The characteristic “box” shape of these distributions arises from SBI’s tendency to bias inferences toward the center of the training distribution, as illustrated by the scatter plots in the upper panels.

\end{document}